\documentclass{aa}

\usepackage{txfonts}
\usepackage{graphicx}
\usepackage{natbib}
\bibpunct{(}{)}{;}{a}{}{,}

\newcommand{\bprp}{G_{\rm BP} - G_{\rm RP}}

\begin{document}

\title{J-PLUS: Towards an homogeneous photometric calibration using {\it Gaia} BP/RP low-resolution spectra}

\author{C.~L\'opez-Sanjuan\inst{1}
\and H.~V\'azquez Rami\'o\inst{1}
\and K.~Xiao\inst{2}
\and H.~Yuan\inst{2}
\and J.~M.~Carrasco\inst{3,4,5}
\and J.~Varela\inst{6}
\and D.~Crist\'obal-Hornillos\inst{6}
\and P.~-E.~Tremblay\inst{7}
\and A.~Ederoclite\inst{1}
\and A.~Mar\'{\i}n-Franch\inst{1}
\and A.~J.~Cenarro\inst{1}
\and P.~R.~T.~Coelho\inst{8}
\and S.~Daflon\inst{9}
\and A.~del Pino\inst{1}
\and H.~Dom\'{\i}nguez S\'anchez\inst{1}
\and J.~A.~Fern\'andez-Ontiveros\inst{1}
\and A.~Hern\'an-Caballero\inst{1}
\and F.~M.~Jim\'enez-Esteban\inst{10}
\and J.~Alcaniz\inst{9}
\and R.~E.~Angulo\inst{11,12}
\and R.~A.~Dupke\inst{9,13,14}
\and C.~Hern\'andez-Monteagudo\inst{15,16}
\and M.~Moles\inst{6}
\and L.~Sodr\'e Jr.\inst{8}
}

\institute{Centro de Estudios de F\'{\i}sica del Cosmos de Arag\'on (CEFCA), Unidad Asociada al CSIC, Plaza San Juan 1, 44001 Teruel, Spain\\\email{clsj@cefca.es}
        \and
        Department of Astronomy, Beijing Normal University, Beijing 100875, People's Republic of China
        \and
        Institut de Ci\`encies del Cosmos (ICCUB), Universitat de Barcelona (UB), Mart\'i i Franqu\`es 1, 08028 Barcelona, Spain
        \and
        Departament de Física Qu\`antica i Astrof\'isica (FQA), Universitat de Barcelona (UB), Mart\'i i Franqu\`es 1, 08028 Barcelona, Spain
        \and
        Institut d'Estudis Espacials de Catalunya (IEEC), c. Gran Capit\`a, 2-4, 08034 Barcelona, Spain
        \and
        Centro de Estudios de F\'{\i}sica del Cosmos de Arag\'on (CEFCA), Plaza San Juan 1, 44001 Teruel, Spain
        \and
        Department of Physics, University of Warwick, Coventry, CV4 7AL, UK
        \and
        Universidade de S\~ao Paulo, Instituto de Astronomia, Geof\'{\i}sica e Ci\^encias Atmosf\'ericas, 05508-090 S\~ao Paulo, Brazil
        \and
        Observat\'orio Nacional - MCTI (ON), Rua Gal Jos\'e Cristino, 77, S\~ao Crist\'ov\~ao, 20921-400 Rio de Janeiro, Brazil
        \and
        Centro de Astrobiología (CAB), CSIC-INTA, Camino Bajo del Castillo s/n, Campus ESAC, 28692 Villanueva de la Ca\~nada, Madrid, Spain
        \and
        Donostia International Physics Centre (DIPC), Paseo Manuel de Lardizabal 4, 20018 Donostia-San Sebastián, Spain
        \and
        IKERBASQUE, Basque Foundation for Science, 48013, Bilbao, Spain
        \and
        University of Michigan, Department of Astronomy, 1085 South University Ave., Ann Arbor, MI 48109, USA
        \and
        University of Alabama, Department of Physics and Astronomy, Gallalee Hall, Tuscaloosa, AL 35401, USA
        \and
        Instituto de Astrof\'{\i}sica de Canarias (IAC), 38205 La Laguna, Spain
        \and
        Departamento de Astrof\'{\i}sica, Universidad de La Laguna (ULL), 38200 La Laguna, Spain
}

\date{Received XX XX 2023 / Accepted XX XX XX}

\abstract
{}
{We present the photometric calibration of the twelve optical passbands for the Javalambre Photometric Local Universe Survey (J-PLUS) third data release (DR3), comprising $1\,642$ pointings of two square degrees each.}
{We selected nearly $1.5$ million main sequence stars with a signal-to-noise ratio larger than ten in the twelve J-PLUS passbands and available low-resolution ($R = 20-80$) spectrum from the blue and red photometers (BP/RP) in {\it Gaia} DR3. We compared the synthetic photometry from BP/RP spectra with the J-PLUS instrumental magnitudes, after correcting for the magnitude and color terms between both systems, to obtain an homogeneous photometric solution for J-PLUS. To circumvent the current limitations in the absolute calibration of the BP/RP spectra, the absolute color scale was derived using the locus of $109$ white dwarfs closer than $100$ pc with a negligible interstellar extinction. Finally, the absolute flux scale was anchored to the Panoramic Survey Telescope and Rapid Response System (Pan-STARRS) photometry in the $r$ band.}
{The precision of the J-PLUS photometric calibration, estimated from duplicated objects observed in adjacent pointings and by comparison with the spectro-photometric standard star GD $153$, is $\sim 12$ mmag in $u$, $J0378$, and $J0395$; and $\sim 7$ mmag in $J0410$, $J0430$, $g$, $J0515$, $r$, $J0660$, $i$, $J0861$, and $z$. The estimated accuracy in the calibration along the surveyed area is better than $1$\% for all the passbands.}
{The {\it Gaia} BP/RP spectra provide a high-quality, homogeneous photometric reference in the optical range across the full-sky, in spite of their current limitations as an absolute reference. The calibration method for J-PLUS DR3 reaches an absolute precision and accuracy of $1$\% in the twelve optical filters within an area of $3\,284$ square degrees.}


\titlerunning{J-PLUS. Towards an homogeneous photometric calibration using {\it Gaia} BP/RP low-resolution spectra}

\authorrunning{L\'opez-Sanjuan et al.}

\maketitle

\section{Introduction}\label{sec:intro}
A fundamental step in the data processing of any imaging survey is its photometric calibration, that translates the observed counts in the reduced images to a physical flux scale referred to the top of the atmosphere. Accurate colors are needed to derive atmospheric parameters for Milky Way stars, photometric redshifts for galaxies and quasars, and surface composition for minor bodies in the Solar System; while reliable absolute fluxes directly affect the estimation of the luminosity and the mass of galaxies and stars. Within this framework, photometric surveys target a calibration uncertainty at the 1\% level and below.

The calibration process can be split in two main steps: obtain a homogeneous photometric solution along the surveyed area, and estimate the absolute flux scale for each passband. Both steps are challenging for large-area (thousand of square degrees) multi-filter (dozens of passbands) surveys, such as the Javalambre Photometric Local Universe Survey (J-PLUS, 12 optical filters; \citealt{cenarro19}), its southern counterpart S-PLUS \citep{splus}, or the Javalambre Physics of the Accelerating Universe Astrophysical Survey (J-PAS, 56 optical filters of $14.5$ nm width; \citealt{jpas, minijpas}). Because of their large number of filters, the observation of spectro-photometric standard stars to perform each night calibration is unfeasible. 

Regarding the homogenization of the photometry, several techniques have been proposed in the literature. We highlight the \"ubercalibration \citep{ubercalsdss, wittman12}, the hypercalibration \citep{finkbeiner16,zhou18}, the forward global modeling \citep{burke18}, the stellar locus regression \citep{covey07, high09, kelly14,clsj19jcal,clsj21zsl}, and the stellar color regression \citep[SCR,][]{scr,huang21,niu21_dr3,xiao22,huang22}. The {\it Gaia} third data release (DR3, \citealt{gaiadr3}) provides for the first time $220$ million low-resolution ($R = 20-80$) spectra \citep{carrasco21,gaiadr3_bprp, gaiadr3_calib} thanks to the observations performed with the blue photometer (BP, $330 - 680$ nm) and the red photometer (RP, $630 - 1\,050$ nm) on board the {\it Gaia} satellite \citep{gaia}. Synthetic photometry from the BP/RP spectra may provide an homogeneous, all-sky, space-based reference for ground-based photometric surveys \citep[][hereafter GC22]{gaiadr3_synphot}, despite the current limitations in the absolute scale of the BP/RP spectra as reflected by the existence magnitude and color terms when compared with well established photometric systems (see GC22, for a detailed discussion). This offers a great opportunity to homogenize the photometry of large-area multi-filter optical surveys, avoiding dedicated observations for calibration and maximizing the survey speed.

In the present paper, we used the {\it Gaia} BP/RP spectra to homogenize the photometric solution of the J-PLUS DR3, covering $3\,284$ deg$^2$ with twelve optical filters (Table~\ref{tab:JPLUS_filters}). The absolute color scale was derived using the white dwarf locus technique presented in \citet{clsj19jcal}. Finally, the absolute flux scale was anchored to the Panoramic Survey Telescope and Rapid Response System (Pan-STARRS) photometry in the $r$ band.

This paper is organized as follows. The J-PLUS DR3 and the ancillary data used are presented in Sect.~\ref{sec:data}. The calibration methodology is summarized in Sect.~\ref{sec:method}, with special emphasis in the use of {\it Gaia} BP/RP spectra. The precision and accuracy in the J-PLUS DR3 calibration are discussed in Sect.~\ref{sect:discussion}. Finally, we present our conclusions in Sect.~\ref{sec:summary}. Magnitudes are given in the AB system \citep{oke83} unless noted otherwise.

\section{Data}\label{sec:data}
\subsection{J-PLUS photometric data}\label{sec:jplus}
J-PLUS\footnote{\url{www.j-plus.es}} is being conducted at the Observatorio Astrof\'{\i}sico de Javalambre (OAJ, \citealt{oaj}) using the 83\,cm Javalambre Auxiliary Survey Telescope (JAST80) and T80Cam, a panoramic camera with a single charge-coupled device (CCD) of 9.2k $\times$ 9.2k pixels that provides a $2\deg^2$ field of view (FoV) with a pixel scale of 0.55$^{\prime\prime}$pix$^{-1}$ \citep{t80cam}. The twelve bands of the J-PLUS filter system are summarized in Table~\ref{tab:JPLUS_filters}. The J-PLUS observational strategy, image reduction, and scientific goals are presented in \citet{cenarro19}.

\begin{table} 
\caption{J-PLUS photometric system.}
\label{tab:JPLUS_filters}
\centering 
        \begin{tabular}{l c c c}
        \hline\hline\rule{0pt}{3ex} 
        Filter $(\mathcal{X})$   & Central wavelength    & FWHM  & $m_{\rm lim}^{\rm DR3}$ \\\rule{0pt}{2ex} 
                &   [nm]                & [nm]           &  [mag]\tablefootmark{a}        \\
        \hline\rule{0pt}{2ex}
        $u$             &348.5  &50.8           &       20.8    \\ 
        $J0378$         &378.5  &16.8           &       20.8    \\ 
        $J0395$         &395.0  &10.0           &       20.8    \\ 
        $J0410$         &410.0  &20.0           &       21.0    \\ 
        $J0430$         &430.0  &20.0           &       21.0    \\ 
        $g$             &480.3  &140.9          &       21.8    \\ 
        $J0515$         &515.0  &20.0           &       21.0    \\ 
        $r$             &625.4  &138.8          &       21.8    \\ 
        $J0660$         &660.0  &13.8           &       21.0    \\ 
        $i$             &766.8  &153.5          &       21.3    \\ 
        $J0861$         &861.0  &40.0           &       20.4    \\ 
        $z$             &911.4  &140.9          &       20.5    \\ 
        \hline 
\end{tabular}
\tablefoot{
\tablefoottext{a} {Limiting magnitude (5$\sigma$, 3 arcsec diameter aperture) of J-PLUS DR3.}
}
\end{table}

The J-PLUS DR3 comprises $1\,642$ pointings ($3\,284$ deg$^2$) observed and reduced in all survey bands. The limiting magnitudes (5$\sigma$, 3$^{\prime\prime}$ aperture) of the DR3 are presented in Table~\ref{tab:JPLUS_filters} for reference. The median point spread function (PSF) full width at half maximum (FWHM) in the DR3 $r$-band images is 1.1$^{\prime\prime}$. Source detection was done in the $r$ band using \texttt{SExtractor} \citep{sextractor}, and the flux measured in the twelve J-PLUS bands at the position of the detected sources using the aperture defined in the $r$-band image. Objects near the borders of the images, close to bright stars or affected by optical artifacts, were masked. This provides a unique high-quality area of $2\,881$ deg$^2$. The DR3 is publicly available at the J-PLUS website\footnote{\url{www.j-plus.es/datareleases/data_release_dr3}} since 13th December 2022. 

We note that the published J-PLUS DR3 photometry already includes all the calibration steps presented in Sect.~\ref{sec:method}. In addition to J-PLUS photometry, ancillary data from {\it Gaia} and Pan-STARRS were used in the calibration process. These datasets are described in the following sections.

\subsection{Gaia DR3}\label{sec:gaia}
The {\it Gaia} spacecraft is mapping the 3D positions and kinematics of a representative fraction of Milky Way stars \citep{gaia}. The mission will ultimately provide astrometry (positions, proper motions, and parallaxes) and optical spectro-photometry for over a billion stars, as well as radial velocity measurements of more than $100$ million stars.

In the present work, we used the {\it Gaia} DR3 \citep{gaiadr3}, which is based on $34$ months of observations. It contains astrometric determinations and provides integrated photometry in three broadbands, namely $G$ ($330 - 1\,050$ nm), $G_{\rm BP}$ ($330 - 680$ nm), and $G_{\rm RP}$ ($630 - 1\,050$ nm), for 1.5 billion sources with $G < 21$. The {\it Gaia} DR3 also contains BP/RP low-resolution ($R = 20-80$) spectra for $220$ million sources with $G < 17.65$ mag and enough transits to ensure a good signal-to-noise ratio ($S/N$) for the data \citep{gaiadr3_bprp,gaiadr3_calib}. These spectra were used to homogenize the J-PLUS photometric solution along the surveyed area (Sect.~\ref{sec:bprp}).

\subsection{Pan-STARRS DR1}\label{sec:ps1}
The Pan-STARRS1 is a 1.8 m optical and near-infrared telescope located on Mount Haleakala, Hawaii. The telescope is equipped with the Gigapixel Camera 1, consisting of an array of $60$ CCD detectors of $4\,800 \times 4\,800$ pixels each. The 3$\pi$ Steradian Survey (hereafter PS1; \citealt{chambers16}) covers the sky at declination $\delta  > -30^{\circ}$ with five filters, $grizy$ \citep{tonry12}.

Astrometry and photometry were extracted by the Pan-STARRS1 Image Processing Pipeline \citep{ps1pipe,ps1cal,ps1phot,ps1pix}. PS1 photometry features a uniform flux calibration, achieving better than 1\% accuracy over the sky \citep{ps1cal,chambers16,xiao22}. The PS1 first data release (DR1) was made public in December $2016$, providing a static-sky catalog and stacked images for the 3$\pi$ Steradian Survey \citep{ps1data}.

Because of its homogeneous depth, excellent internal calibration, and large footprint; PS1 photometry provides an ideal reference to set the absolute flux scale of the J-PLUS magnitudes (Sect.~\ref{sec:deltar}).

\section{Photometric calibration of J-PLUS DR3}\label{sec:method}
The goal of the calibration process is to obtain the zero point (ZP) of the observation; that relates the magnitude of the sources in a certain passband $\mathcal{X}$ at the top of the atmosphere with the magnitudes obtained from the analogue-to-digital unit (ADU) counts of the reduced images. We simplify the notation in the following using the passband name as the magnitude in such filter. Thus,
\begin{equation}
\mathcal{X} = -2.5\log_{10}({\rm ADU}_{\mathcal{X}}) + {\rm ZP}_{\mathcal{X}}.
\end{equation}
In the estimation of the J-PLUS DR3 instrumental photometry, the reduced images were normalized to a one-second exposure and an arbitrary zero point ${\rm ZP}_{\mathcal{X}} = 25$ was applied. This defined the instrumental magnitudes $\mathcal{X}_{\rm ins}$.

\begin{figure}[t]
\centering
\resizebox{\hsize}{!}{\includegraphics{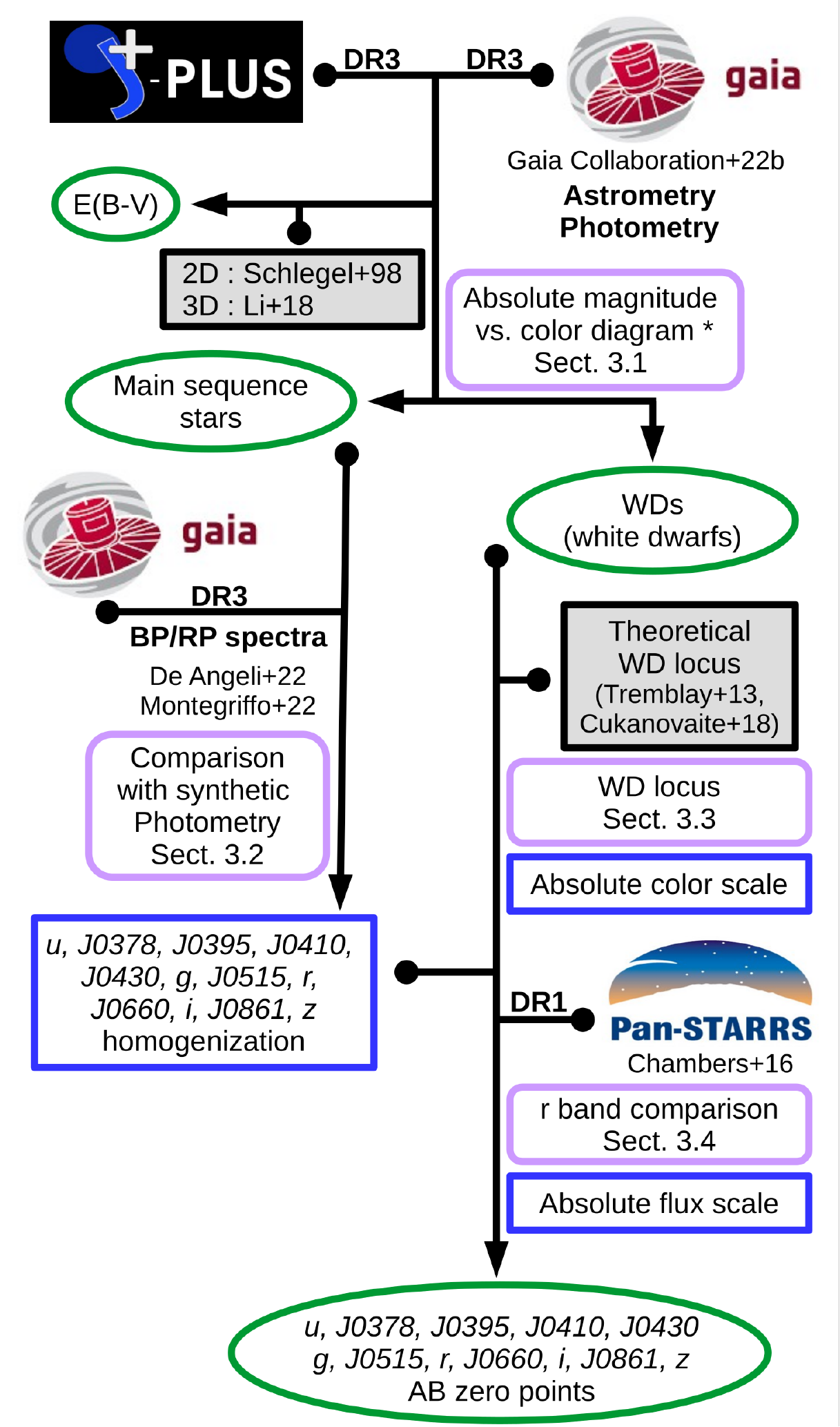}}
\caption{Updated flowchart of the calibration method used in J-PLUS DR3. Arrows that originate in small dots indicate that the preceding data product is an input to the subsequent analysis. Datasets are shown with their project logo, and external data or models are denoted with black boxes. The rounded purple boxes show the calibration steps. The asterisk indicates the step based on dust de-reddened magnitudes. The blue boxes show intermediate data products, and green ovals highlight data products of the calibration process. The main change with respect to J-PLUS DR2 calibration is the use of {\it Gaia} BP/RP low-resolution spectra in the homogenization (Sect.~\ref{sec:bprp}).} 
\label{fig:calib_chart}
\end{figure}

The calibration process applied in J-PLUS DR3 has different steps, as described in the following sections and summarized in Fig.~\ref{fig:calib_chart}. The final outcome is the zero point of the passband $\mathcal{X}$ estimated for the pointing $p_{\rm id}$ as
\begin{align}
{\rm ZP} & _{\mathcal{X}}\,(p_{\rm id},X,Y) = \nonumber \\ & \Delta \mathcal{X}_{\rm atm}\,(p_{\rm id}) + P_{\mathcal{X}}\,(p_{\rm id},X,Y) + \Delta \mathcal{X}_{\rm WD} + \Delta r_{\rm PS1} + 25,\label{eq:zp}
\end{align}
where $\Delta \mathcal{X}_{\rm atm}$ accounts for the atmospheric extinction at the moment of the observation (Sect.~\ref{sec:bprp}), $P_{\mathcal{X}}$ defines a plane that accounts for the 2D variation of the calibration with the $(X,Y)$ position of the sources on the CCD (Sect.~\ref{sec:bprp}), $\Delta \mathcal{X}_{\rm WD}$ is the global offset provided by the white dwarf (WD) locus that translates homogenized colors to the AB scale (Sect.~\ref{sec:wdlocus}), and $\Delta r_{\rm PS1}$ is the global offset for the $r$ band to anchor the absolute flux scale to the PS1 photometric solution (Sect.~\ref{sec:deltar}).

The J-PLUS instrumental magnitudes used for calibration were measured on a $6$ arcsec diameter aperture corrected by aperture effects to retrieve the total flux of stars. The aperture correction $C_{\rm aper}$ depends on the passband and the pointing, and was computed from the growth curve of non-saturated, bright stars in each image. The median aperture correction among all the passbands is $C_{\rm aper} = -0.09$ mag. The corrections used are available in the J-PLUS database\footnote{Column \texttt{APER\_COR\_6\_0} in the table \texttt{jplus.TileImage}.} and additional details about their estimation can be found in \citet{clsj19jcal}.

\begin{figure*}
\sidecaption
\includegraphics[width=12cm]{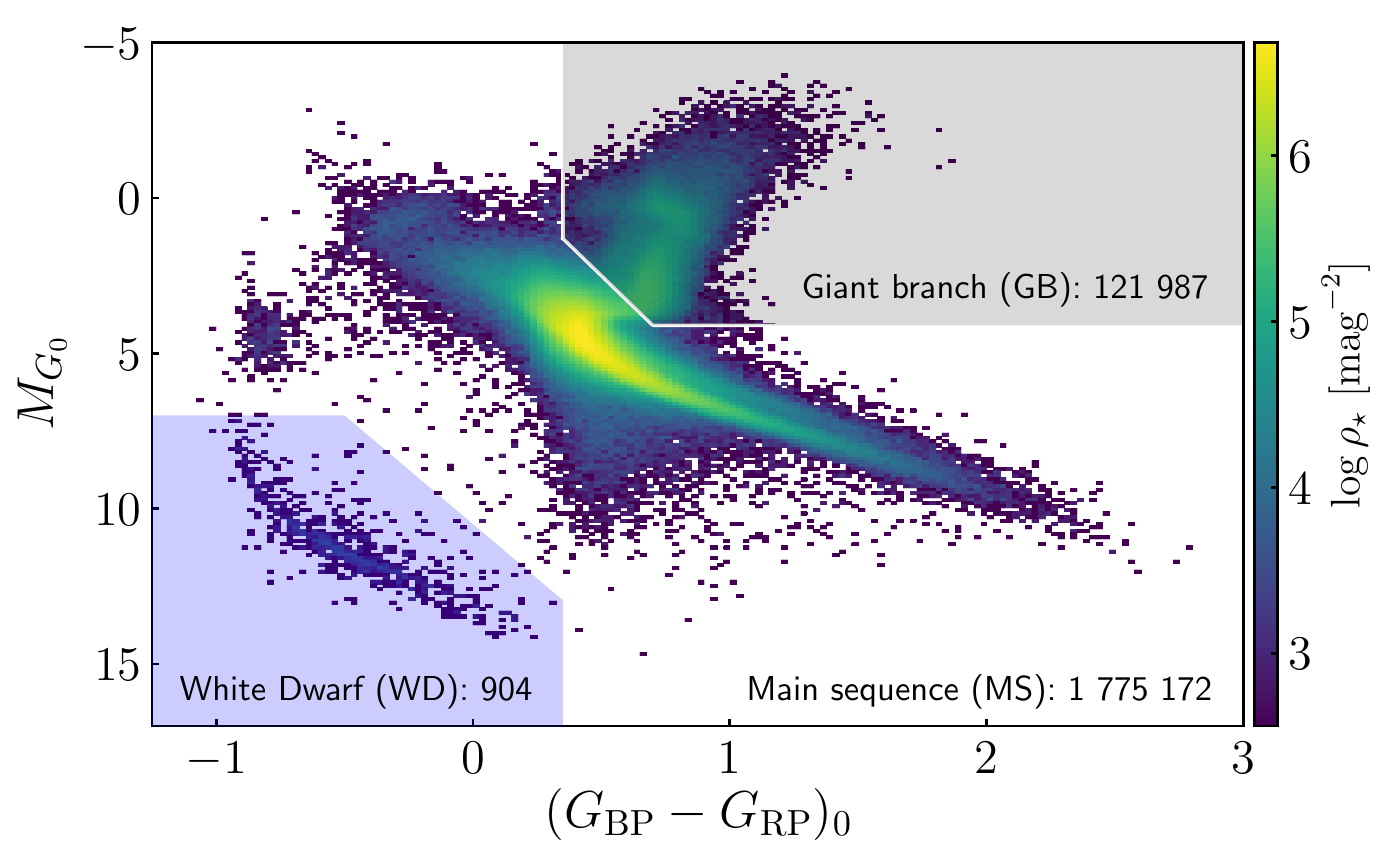}
\centering
\caption{
Absolute magnitude in the $G$ band vs.
$G_{\rm BP} - G_{\rm RP}$ color diagram, corrected for dust reddening, of the $1\,898\,063$ high-quality sources in common between {\it Gaia} DR3 and J-PLUS DR3. The color scale presents the number density of stars per mag$^{2}$, noted $\rho_{\star}$. Three areas were defined following \citet{clsj19jcal}, dominated by main sequence stars (white area), giant branch stars (gray area), and white dwarfs (blue area).} 
\label{fig:hrgaia}
\end{figure*}

\subsection{Step 1: Selection of the calibration stars}\label{sec:stars}
The first step of our methodology is to define a high-quality sample of stars to perform the photometric calibration. We cross-matched the J-PLUS DR3 sources with $S/N > 10$ and \texttt{SExtractor} photometric flag equal to zero (i.e. with neither close detections nor image problems) in all twelve passbands against the {\it Gaia} DR3 catalog using a 1.5$\arcsec$ radius\footnote{The complete J-PLUS DR3 versus {\it Gaia} catalog can be found in the table \texttt{jplus.xmatch\_gaia\_dr3} at the J-PLUS database.}. We retained {\it Gaia} sources with $S/N > 3$ in parallax, represented as $\varpi$ [arcsec], and with a photometric measurement in $G$, $G_{\rm BP}$, and $G_{\rm RP}$. Finally, J-PLUS sources with more than one {\it Gaia} counterpart were discarded. We obtained $1\,898\,063$ unique high-quality stars for calibration.

Then, the {\it Gaia} absolute magnitude versus color diagram was constructed, as presented in Fig.~\ref{fig:hrgaia}. The dust de-reddened $G$ absolute magnitude of the calibration stars was obtained as
\begin{equation}
M_{G_0} = G - k_G E(B-V) + 5\log_{10} \varpi + 5,\label{eq:mg0}
\end{equation}
where $E(B-V)$ is the color excess of the source and $k_{G}$ the extinction coefficient of the $G$ passband. The de-reddened $\bprp$ color was computed as
\begin{equation}
(\bprp)_0 = G_{\rm BP} - k_{G_{\rm BP}} E(B-V) - G_{\rm RP} + k_{G_{\rm RP}} E(B-V),
\end{equation}
where $k_{G_{\rm BP}}$ and $k_{G_{\rm RP}}$ are the extinction coefficients in the $G_{\rm BP}$ and $G_{\rm RP}$ passbands, respectively. The extinction coefficients were obtained with the extinction law presented in \citet{schlafly16} and assuming $R_V = 3.1$; with $k_G = 2.600$, $k_{G_{\rm BP}} = 3.410$, and $k_{G_{\rm RP}} = 1.807$. This provides a first-order correction, since the proper coefficients depends on color and dust-column density \citep[e.g.][]{danielski18,zhang22}.

The color excess at infinite distance of each J-PLUS source was estimated from the \citet{sfd98} extinction map\footnote{Stored in table \texttt{jplus.MWExtinction} at the J-PLUS database.}. The calibration stars have distance information from {\it Gaia} DR3 parallaxes, and we used the Milky Way dust model presented in \cite{li18} to properly scale the color excess at infinity to obtain $E(B-V)$. This process was tested with the star-pair method presented in \citet{yuan13}. We concluded that the assumed $E(B-V)$ is a good proxy for the real color excess of the stars with an uncertainty of $0.012$ mag. Additional details are presented in \citet{clsj21zsl}.

The inverse of the parallax was used as a distance proxy in Eq.~(\ref{eq:mg0}). This is a crude approximation to the distance, as demonstrated by \citet{bailerjones18}. Since our goal is to define general populations to calibrate the J-PLUS photometry, the simplified extinction and distance schemes used in Eq.~(\ref{eq:mg0}) fulfill our requirements.

Following \citet{clsj19jcal}, three areas were defined in the magnitude-color diagram. These areas are dominated by main sequence stars ($1\,775\,172$ sources), giant branch stars ($121\,987$ sources), and white dwarfs ($904$ sources). The main sequence stars were used in the homogenization step (Sect.~\ref{sec:bprp}) and the white dwarfs to obtain the AB scale of the J-PLUS colors (Sect.~\ref{sec:wdlocus}).

\begin{figure}[t]
\centering
\resizebox{\hsize}{!}{\includegraphics{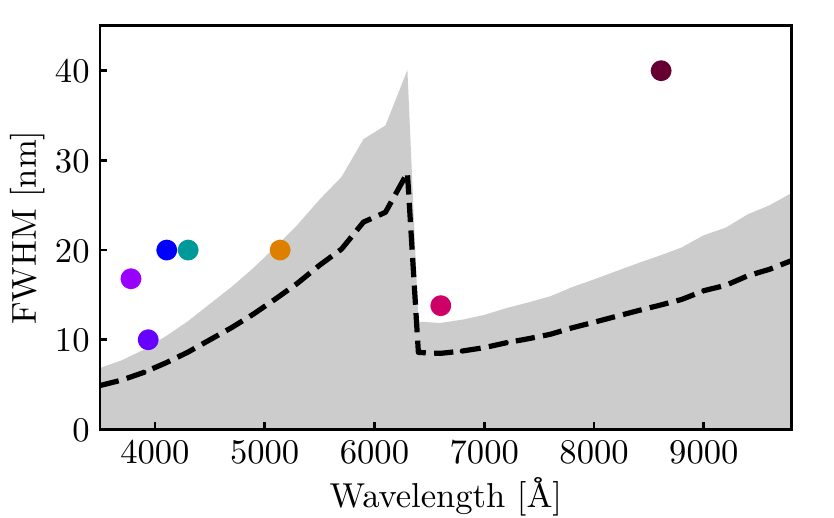}}
\caption{Full width at half maximum of the {\it Gaia} BP/RP spectra as a function of wavelength (black dashed line). The gray area marks the avoidance region corresponding to less than $1.4$ times the FWHM of the BP/RP spectra. The colored dots show the FWHM of the medium and narrow passbands in J-PLUS.
} 
\label{fig:fwhm}
\end{figure}

\begin{figure*}[t]
\centering
\resizebox{\hsize}{!}{\includegraphics{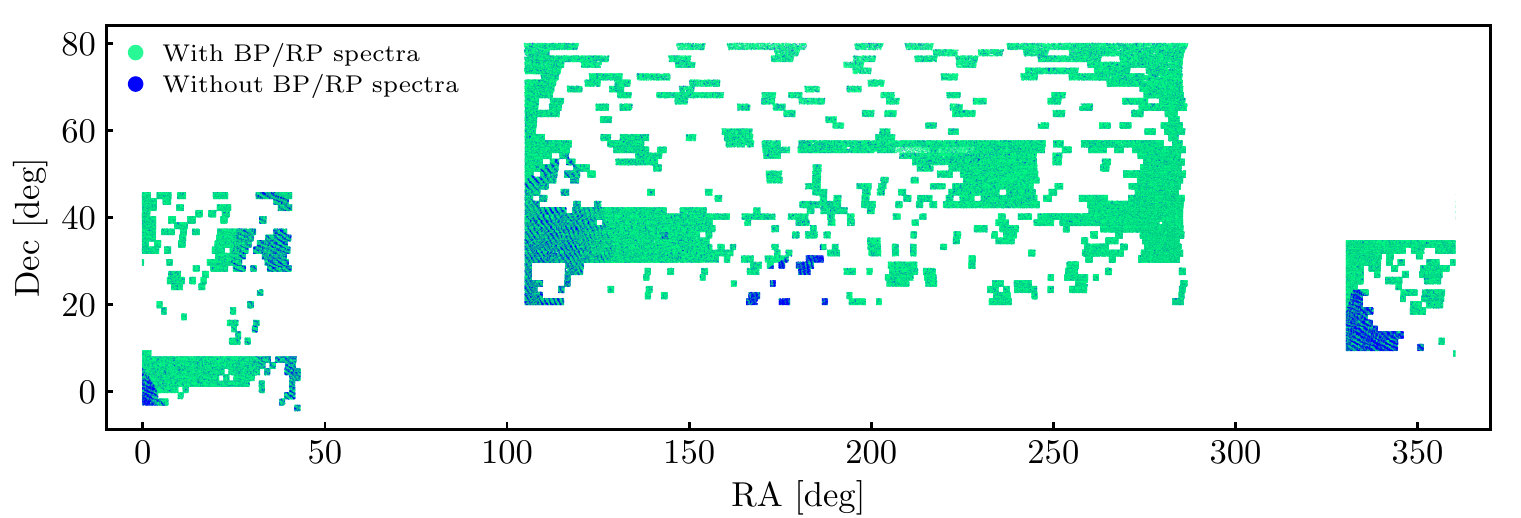}}\\
\resizebox{0.49\hsize}{!}{\includegraphics{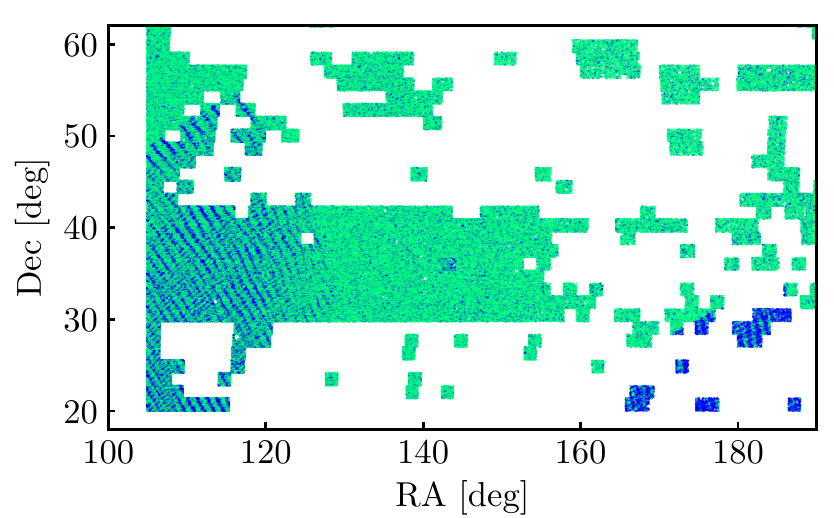}}
\resizebox{0.49\hsize}{!}{\includegraphics{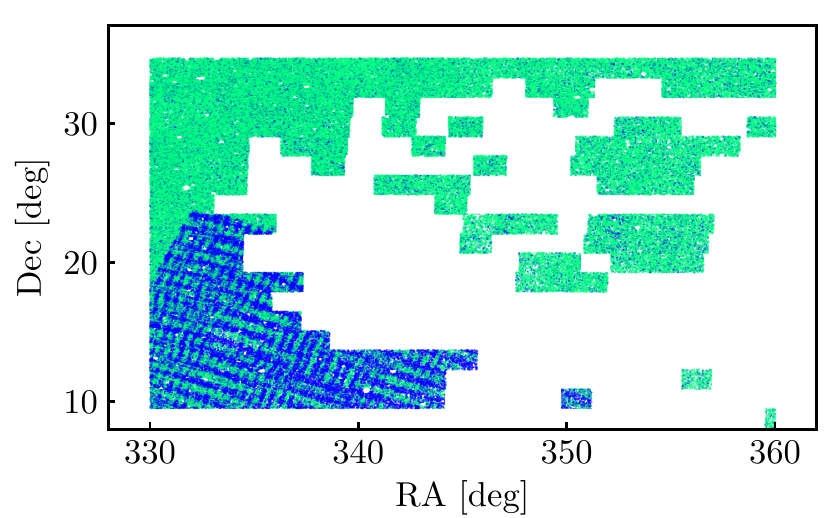}}
\caption{Sky distribution of the main sequence calibration stars with (green dots) and without (blue dots) BP/RP spectra in {\it Gaia} DR3 after applying the recommended quality selection criteria. {\it Top panel}: Full J-PLUS DR3 footprint. {\it Bottom panels}: Zoom in two areas specially affected by the {\it Gaia} scanning law.
} 
\label{fig:gaia_nstar}
\end{figure*}

\subsection{Step 2: Homogenization with synthetic photometry from BP/RP spectra}\label{sec:bprp}
The main change with respect to the calibration process of previous J-PLUS data releases is the replacement of the stellar locus technique by the synthetic photometry from BP/RP spectra to homogenize the photometric solution along the surveyed area. The stellar locus technique demands a previous knowledge of the extinction and some atmospheric parameters of the stars (i.e., surface gravity and metallicity) to avoid systematics across the sky \citep{clsj19jcal,clsj21zsl}. The stellar locus technique is therefore limited by our current understanding of the interstellar extinction \citep[e.g.][]{sun22} and the access to spectroscopic-based metallicities. The {\it Gaia} BP/RP low-resolution spectra provide a great opportunity to have an all-sky, space-based reference photometry and obtain an homogeneous J-PLUS calibration without the need of previous knowledge about the extinction or the metallicity of the used stars. Nevertheless, we note that any systematics present in the BP/RP spectra will be inherited by J-PLUS.

\begin{figure}[t]
\centering
\resizebox{\hsize}{!}{\includegraphics{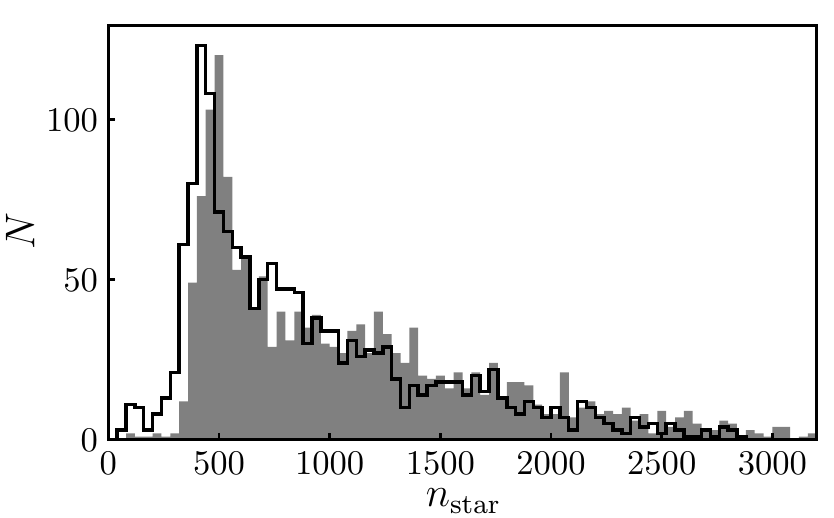}}
\caption{Histogram of J-PLUS DR3 pointings for a given number of main sequence calibration stars ($n_{\rm star}$) with (black empty) and without (gray filled) BP/RP spectra in {\it Gaia} DR3 after applying the recommended quality selection cuts.} 
\label{fig:nstar_hist}
\end{figure}

\begin{figure*}[th]
\centering
\resizebox{0.49\hsize}{!}{\includegraphics{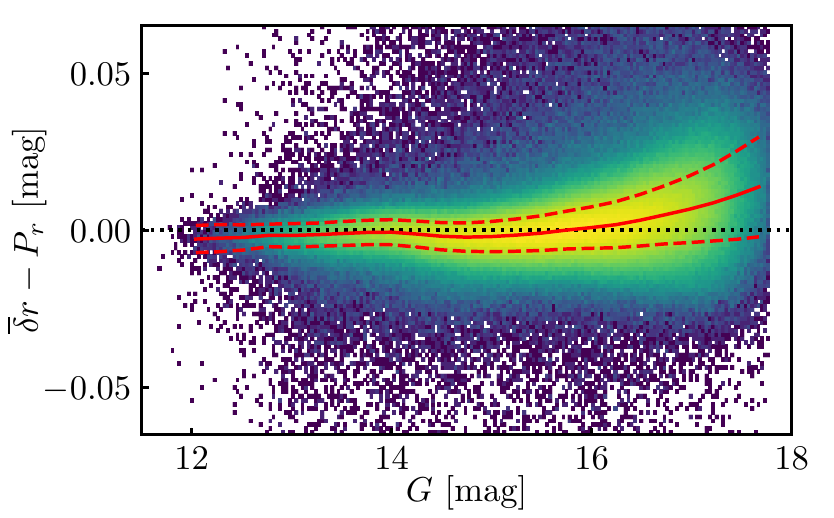}}
\resizebox{0.49\hsize}{!}{\includegraphics{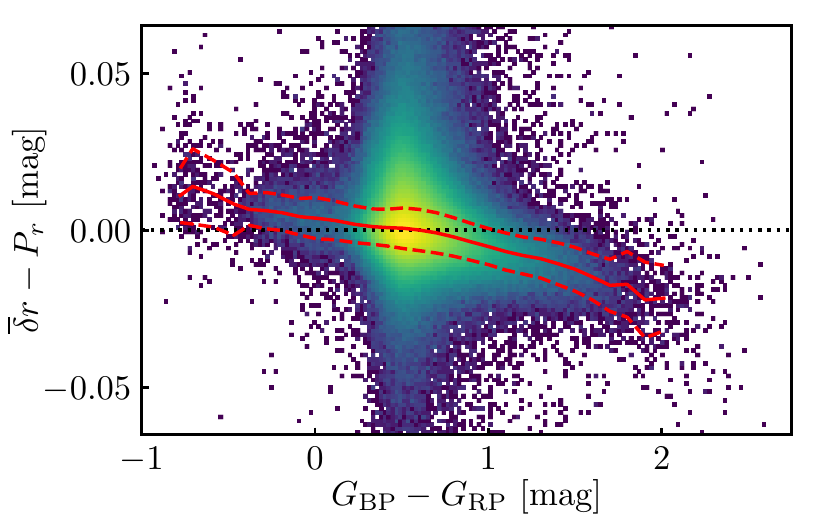}}\\
\resizebox{0.49\hsize}{!}{\includegraphics{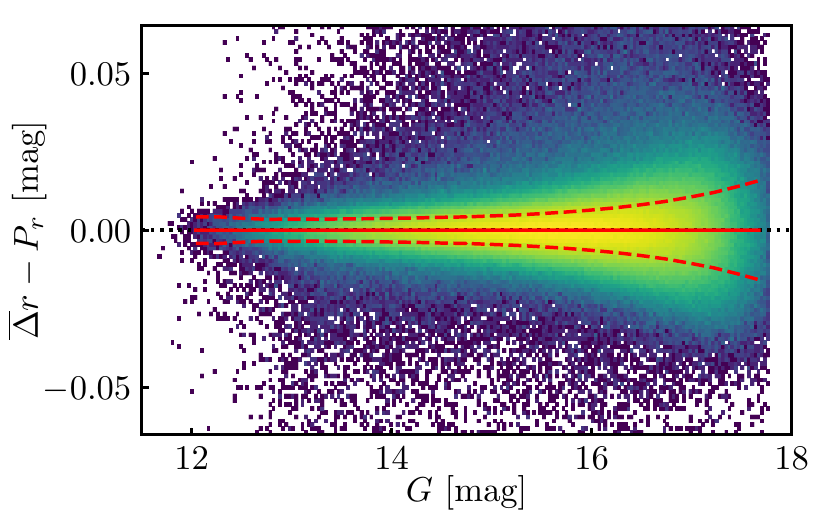}}
\resizebox{0.49\hsize}{!}{\includegraphics{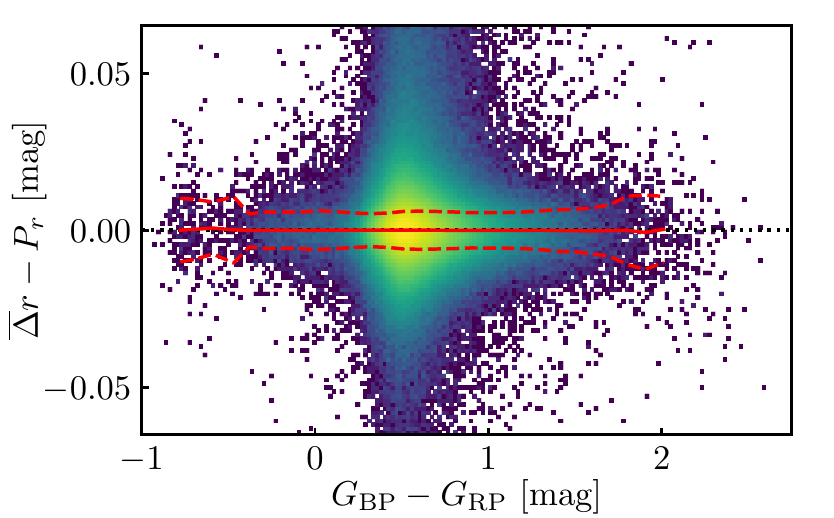}}\\
\caption{Residuals between the synthetic photometry from {\it Gaia} BP/RP spectra and J-PLUS photometry in the $r$ band as a function of the $G$ magnitude (left panels) and the $\bprp$ color (right panels) without the transformation terms $T_{r}^{\rm mag}$ and $T_{r}^{\rm col}$ (top panels) and after applying them (bottom panels). The color scale depicts the number density of sources using a logarithm scale, with nearly $1.5$ million sources shown. In all the panels, the solid-red line represents the the median of the residuals and the dashed lines its one $\sigma$ dispersion. 
} 
\label{fig:Tr}
\end{figure*}

\begin{figure*}[th]
\centering
\resizebox{0.49\hsize}{!}{\includegraphics{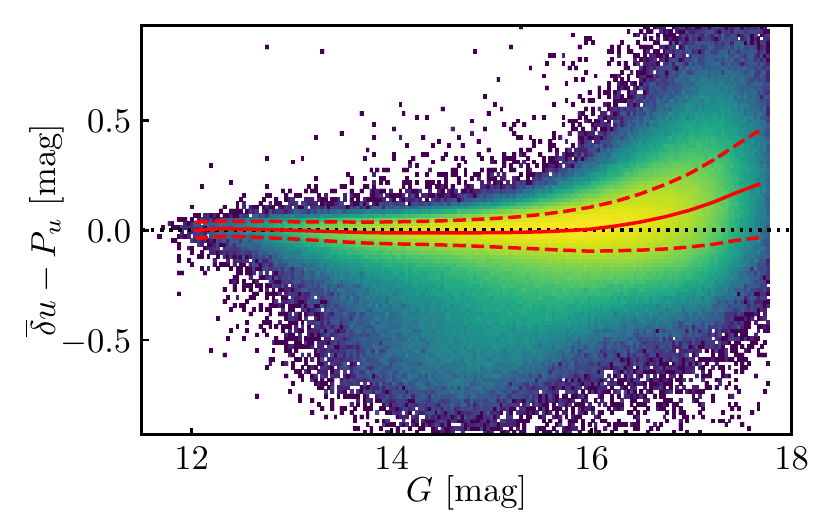}}
\resizebox{0.49\hsize}{!}{\includegraphics{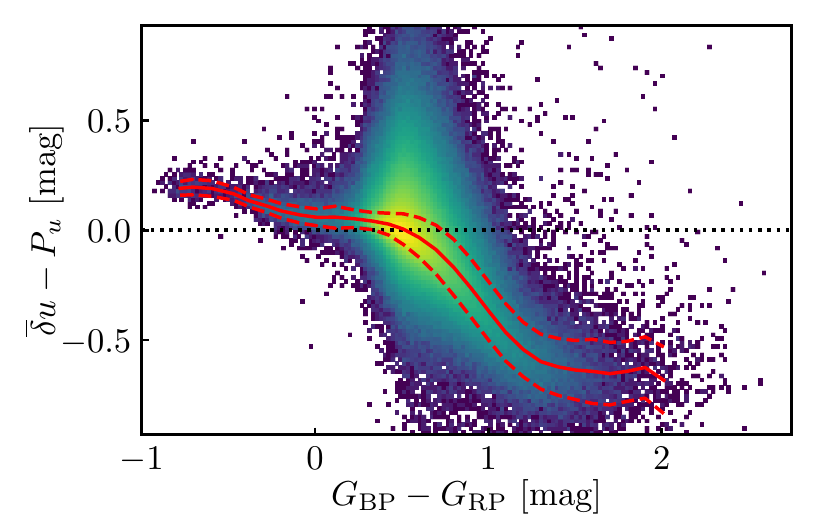}}\\
\resizebox{0.49\hsize}{!}{\includegraphics{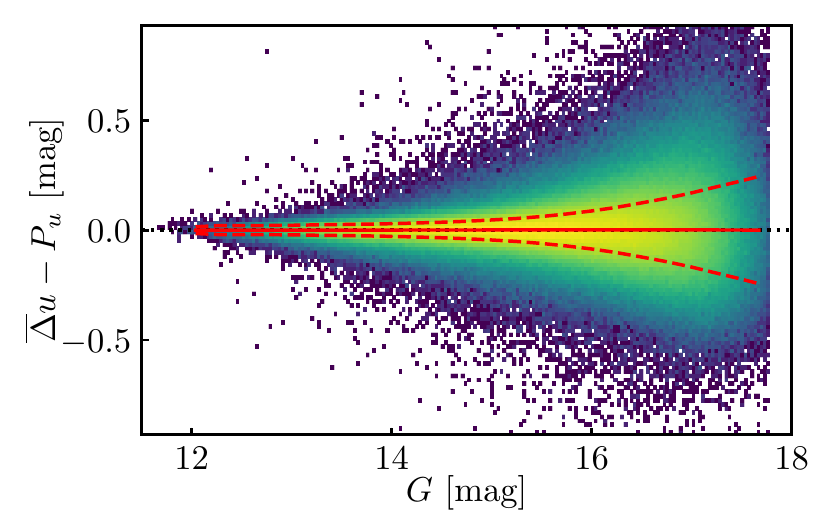}}
\resizebox{0.49\hsize}{!}{\includegraphics{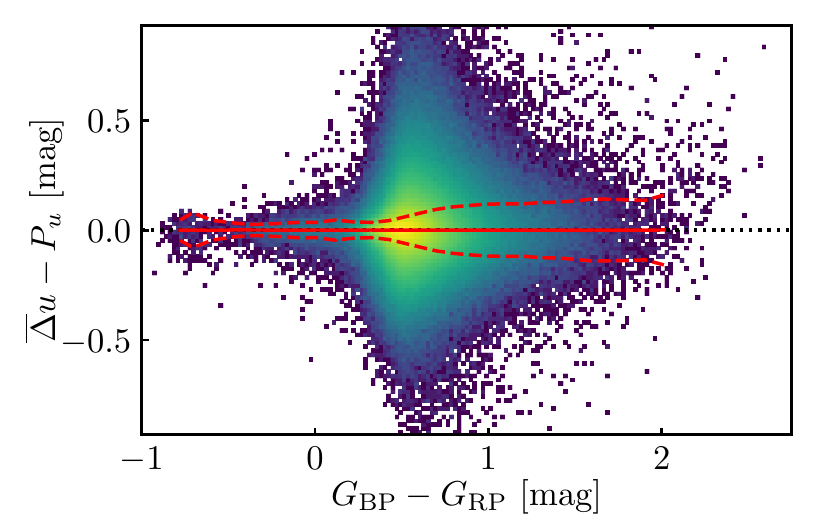}}\\
\caption{Residuals between the synthetic photometry from {\it Gaia} BP/RP spectra and J-PLUS photometry in the $u$ band, following Fig.~\ref{fig:Tr}.
} 
\label{fig:Tu}
\end{figure*}

The basis of the synthetic photometry estimation for a given passband from the BP/RP spectra are extensively presented in GC22. We follow their recommendations and suggestions to obtain the proper synthetic photometry in the J-PLUS filter system. The first goal is to ensure that the synthetic photometry derived from {\it Gaia} BP/RP spectra is reliable. We followed the recipe in GC22, where the FWHM of the targeted passband should be at least $1.4$ times larger than the FWHM of the externally calibrated {\it Gaia} spectra \citep{gaiadr3_calib} at the central wavelength of the passband. The comparison between the FWHM of the J-PLUS medium and narrow passbands and of the BP/RP spectra is presented in Fig.~\ref{fig:fwhm}. We found that the J-PLUS filter system can be safely obtained from BP/RP spectra. The FWHM of all the passbands is larger than the 1.4-times limit, with $J0395$ and $J0515$ just on the edge.

The next stage was to obtain the synthetic photometry of the main sequence calibration stars. We used the code \texttt{GaiaXPy}\footnote{\url{https://gaia-dpci.github.io/GaiaXPy-website}} to retrieve the J-PLUS synthetic magnitudes, noted $\mathcal{X}_{\rm syn}$, for those sources with available spectrum and $G < 17.65$ Vega mag. In the process, additional quality cuts were applied: $\texttt{RUWE} < 1.4$, $\texttt{phot\_variable\_flag} \neq {\rm VARIABLE}$, and the 5$\sigma$ condition for the flux excess factor $C^{*}$ defined by \citet{riello21}. We refer the reader to the {\it Gaia} documentation and the work by \citet{riello21} for the definition of these fields. To minimize the impact of blended sources in the J-PLUS photometry, only those with a morphological $\texttt{class\_star} > 0.1$ in the J-PLUS database were considered. The final number of sources with synthetic photometry from BP/RP spectra was $1\,498\,074$. The distribution of these sources in the sky is presented in Fig.~\ref{fig:gaia_nstar}. We found that most J-PLUS pointings present an homogeneous coverage, but some areas are underpopulated. This is a consequence of the {\it Gaia} scanning law and the lower number of observations available in the missing areas. As already mentioned by GC22, these areas will be filled in future {\it Gaia} data releases. We checked the impact of the depopulated areas in the number of calibration stars per pointing, $n_{\rm star}$ (Fig.~\ref{fig:nstar_hist}). We found that the median number of calibration sources is $n_{\rm star} = 751$, with a mode of $430$ sources. The median fraction of missing calibration stars per pointing with respect to the initial sample is $11$\%. There are only $27$ ($1.6$\%) pointings heavily affected by the missing areas and less than $200$ calibration sources. We tested that the general calibration process worked correctly even in these pointings, and therefore we did not apply any further correction to them.

\begin{figure*}[t]
\centering
\resizebox{0.49\hsize}{!}{\includegraphics{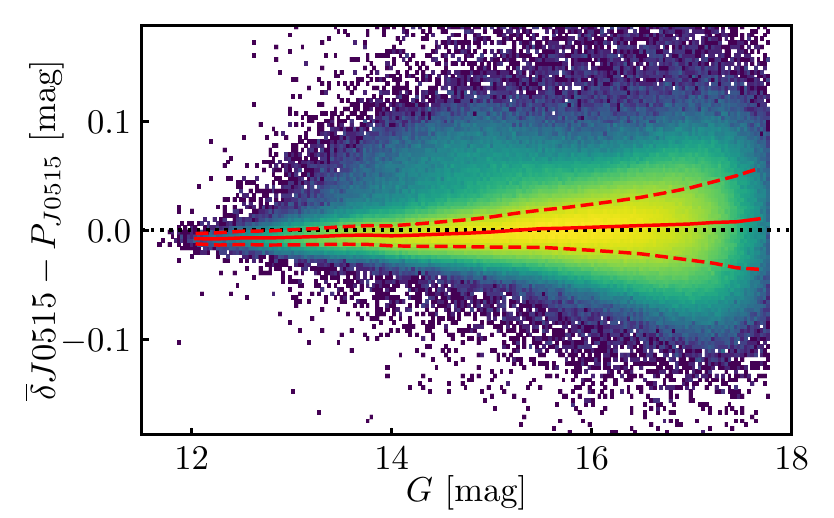}}
\resizebox{0.49\hsize}{!}{\includegraphics{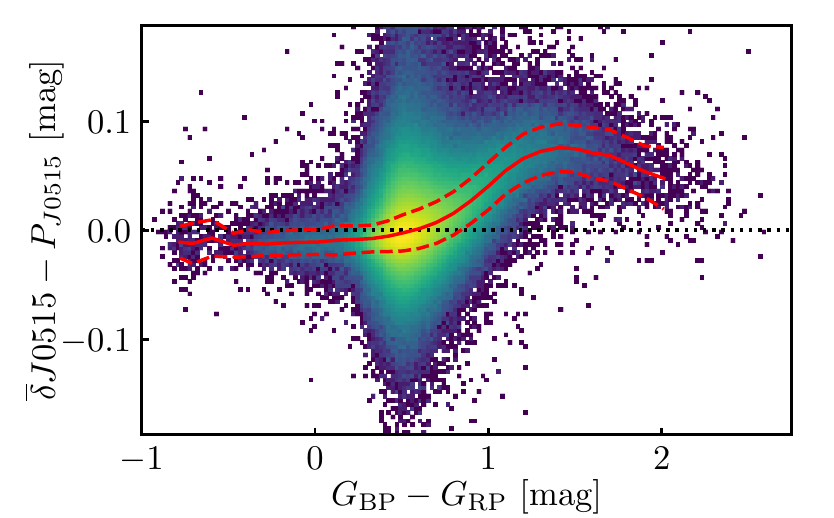}}\\
\resizebox{0.49\hsize}{!}{\includegraphics{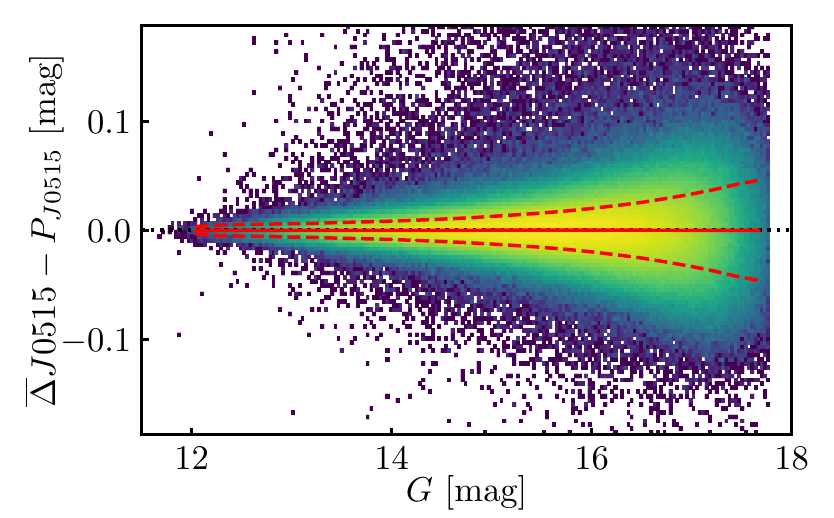}}
\resizebox{0.49\hsize}{!}{\includegraphics{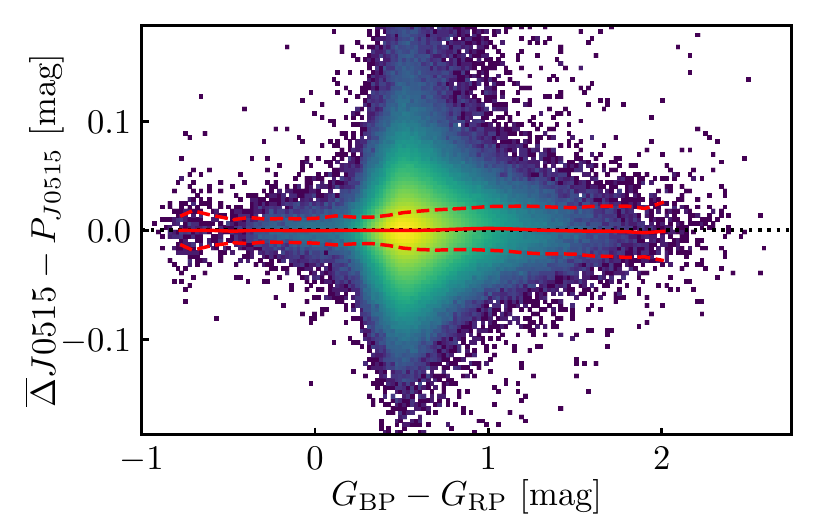}}\\
\caption{Residuals between the synthetic photometry from {\it Gaia} BP/RP spectra and J-PLUS photometry in the $J0515$ band, following Fig.~\ref{fig:Tr}.
} 
\label{fig:Tj0515}
\end{figure*}

We measured the difference between the J-PLUS synthetic photometry from BP/RP spectra and the J-PLUS instrumental photometry for each star in a given pointing $p_{\rm id}$ and filter as
\begin{equation}
    \delta \mathcal{X} = \mathcal{X}_{\rm syn} - \mathcal{X}_{\rm ins}.
\end{equation}
The distribution $\delta \mathcal{X}$ was fitted with a Gaussian function of median $\mu_{\mathcal{X}}$ and dispersion $\sigma_{\mathcal{X}}$. Then, the zero-median difference for each star was defined as
\begin{equation}
    \overline{\delta} \mathcal{X} = \mathcal{X}_{\rm syn} - \mathcal{X}_{\rm ins} - \mu_{\mathcal{X}}.
\end{equation}
As shown by \citet{clsj19jcal}, the residuals $\overline{\delta} \mathcal{X}$ vary along the FoV and are position dependent. The dominant spatial component has a plane shape, and we performed a fit to the function
\begin{equation}
    P_{\mathcal{X}}\,(p_{\rm id},X,Y) = A \cdot X + B \cdot Y + C,
\end{equation}
where $(X,Y)$ represents the location of the source in the CCD\footnote{Variables \texttt{X\_IMAGE} and \texttt{Y\_IMAGE} on the J-PLUS database.}. We applied the plane correction and re-evaluated the median of the distribution to obtain the final residuals.

This process should provide the term $\Delta \mathcal{X}_{\rm atm}$ in Eq.~(\ref{eq:zp}). However, GC22 demonstrate that the current absolute calibration of the BP/RP spectra has both magnitude and color terms when compared with well-established observations in a variety of photometric systems (i.e. Sloan Digital Sky Survey, Pan-STARRS, Johnson-Kron-Cousins). We found that these terms are also present in J-PLUS, as shown in the top panels of Figs.~\ref{fig:Tr}, \ref{fig:Tu}, and \ref{fig:Tj0515} for the $r$, $u$, and $J0515$ passbands, respectively. Similar figures for the remaining J-PLUS filters are presented in Appendix~\ref{app:tfilter}. We note that the definition of our residuals have an opposite sign to the definition in GC22. On the one hand, the hockey stick feature found by \cite{evans18}, \citet{riello21}, and GC22 in the magnitude residuals is also clear in J-PLUS. On the other hand, the color terms are also consistent with the findings from GC22 for similar broad bands. We found that the color terms in the bluest passbands are large. For example, the $u$ band present a difference of $0.2$ mag at $\bprp \sim -1$ mag, and $-0.6$ mag at $\bprp \sim 2$ mag (Fig.~\ref{fig:Tu}). We note that even in this extreme case, the differences are systematic and nearly independent of the J-PLUS pointing. Finally, the medium and narrow passbands have different behaviors. As an example, the $J0515$ passband presents a small difference at $\bprp < 0.5$ mag, that increases to reach $0.1$ mag at $\bprp \sim 1.5$ mag (Fig.~\ref{fig:Tj0515}). 

From the measured magnitude and color differences, we estimated transformation functions to translate as accurately as possible $\mathcal{X}_{\rm syn}$ to $\mathcal{X}_{\rm ins}$. These functions are noted $T_{\mathcal{X}}^{\rm mag}$ and $T_{\mathcal{X}}^{\rm col}$ for the magnitude and color terms, respectively. To compute them, the median of the residuals was evaluated for $G \in [11.75,18.00]$ Vega mag in 0.25 mag bins and $G_{\rm BP} - G_{\rm RP} \in [-0.5,2.5]$ Vega mag in 0.1 mag bins. Then, a linear interpolation in magnitude and color was done independently. For those sources beyond the magnitude or color limits, that correspond to only $0.01$\% of the calibration stars, a linear extrapolation was applied. We defined the transformed differences for each star as
\begin{equation}
    \Delta \mathcal{X} = \mathcal{X}_{\rm syn} - T_{\mathcal{X}}^{\rm mag} - T_{\mathcal{X}}^{\rm col} - \mathcal{X}_{\rm ins},
\end{equation}
and updated the value of $\mu_{\mathcal{X}}$. The zero-median difference, $\overline{\Delta} \mathcal{X} = \Delta \mathcal{X} - \mu_{\mathcal{X}}$, was used to re-evaluate the plane correction and the new residuals to update the transformation functions. This process was iterated five times, converging to median magnitude and color terms below $1$ mmag (bottom panels in Figs.~\ref{fig:Tr}, \ref{fig:Tu}, and \ref{fig:Tj0515}).

During the estimation of the calibration accuracy across the surveyed area (Sect.~\ref{sec:accuracy}), we found systematic discrepancies between the {\it Gaia}-based zero points and those measured with the SCR method in those pointings more affected by interstellar reddening. To account for this fact, in the estimation of the final differences the function $T_{\mathcal{X}}^{\rm col}$ was evaluated at
\begin{equation}
    (\bprp)^{*} = (\bprp) - \delta (\bprp)
\end{equation}
for those pointings with $\delta (\bprp) > 0.015$ mag, where $\delta (\bprp)$ is the difference in the median color of the calibration stars in the pointing with respect to the median for all the calibration sources, $\langle \bprp \rangle = 0.887$ Vega mag. This process implies that the shape of the color transformation is equivalent between pointings, but significantly displaced in those areas with a relevant interstellar extinction. We mitigated the impact of this issue by matching the median of the colors in those pointings with a significant reddening. The J-PLUS footprint does not cover heavily extincted areas, and thus the adopted hypothesis should be tested in the future using either different datasets or archival JAST80 observations performed for open time programs covering highly extincted regions.

The median in the final distribution of $\Delta \mathcal{X} - P_{\mathcal{X}}$ was stored as the term $\Delta \mathcal{X}_{\rm atm}$ for each passband and pointing. At this stage, we defined the homogenized J-PLUS magnitudes from {\it Gaia} BP/RP spectra as
\begin{equation}
    \mathcal{X}_{\rm G} = \mathcal{X}_{\rm ins} +  \Delta \mathcal{X}_{\rm atm} + P_{\mathcal{X}}.
\end{equation}
The relative precision and accuracy of these magnitudes are analyzed in Sect.~\ref{sect:discussion}. 


\subsection{Step 3: Absolute color scale with the white dwarf locus}\label{sec:wdlocus}
The homogeneous J-PLUS magnitudes derived in the previous section must be translated from the {\it Gaia} scale to the AB scale. As shown by GC22 for several filter systems and in the present paper for J-PLUS, the presence of magnitude and color terms are due to the current limitations in the external, absolute calibration of {\it Gaia} DR3 BP/RP spectra. To circumvent this problem, the absolute color scale of the J-PLUS passbands was obtained with the white dwarf locus technique. Here, we provide the relevant technical details for completeness, and the reader is referred to \citet{clsj19jcal} for a detailed description.

The properties of white dwarfs make them excellent standard sources for calibration \citep{holberg06, wall19}. Their model atmospheres can be specified at the $\approx 1$\% level with the knowledge of the effective temperature ($T_{\rm eff}$) and the surface gravity ($\log {\rm g}$). These parameters can be estimated from spectroscopy, providing a reference flux for calibration. They are also mostly photometrically stable. A significant theoretical and observational effort is still underway to provide a robust white dwarf network for the calibration of photometric surveys \citep[e.g.][and references therein]{bohlin00,holberg06,narayan16,narayan19,GF20,bohlin20}.

The observational white dwarf locus presents two branches, corresponding to hydrogen- and helium-dominated atmospheres \citep[e.g.,][]{holberg06,ivezic07,cfisu,gentilefusillo19,bergeron19,clsj22pda}. We performed a Bayesian modeling of the eleven independent $(\mathcal{X} - r)_{\rm G}$ versus $(g-i)_{\rm G}$ color-color diagrams in J-PLUS, with the $r$ band used as the absolute reference in the process. We confronted the theoretical locus against the observations, accounting for the observational errors in the colors, and estimated the best parameters that model the observed color-color distribution of the white dwarfs. The parameter space was explored with the Monte Carlo Markov chain code \texttt{emcee} \citep{emcee}.

The theoretical loci for hydrogen- and helium-dominated atmospheres were obtained from the models including 3D effects presented in \citet[][pure-H atmospheres]{tremblay13} and \citet[][pure-He atmospheres]{cukanovaite18}, respectively. The locus model has $26$ parameters. The distribution in $(g-i)_{\rm G}$ was described by a Gaussian function, whose parameters were the median ($\mu$) and the dispersion ($s$). The general white dwarf population has two parameters: the fraction of H-dominated white dwarfs ($f_{\rm H}$) and the median surface gravity of the population. The offsets in each color-color diagram account for eleven parameters, named $\Delta \mathcal{C}_1$ and $\Delta \mathcal{C}_2$. These offsets impose a match between the theoretical locus and the observations. The offset $\Delta \mathcal{C}_2$ is equivalent to $-\Delta \mathcal{X}_{\rm WD}$ in Eq.~(\ref{eq:zp}), translating the homogenized photometry to the AB scale. We defined $\Delta \mathcal{C}_1 = \Delta i_{\rm WD} - \Delta g_{\rm WD}$, a term shared by all the color-color diagrams. This reduced the initial $22$ parameters to eleven independent measurements. Finally, the diversity of white dwarf properties produces a physical dispersion in the locus after accounting for observational uncertainties. These physical variations are encoded in an intrinsic dispersion for each passband ($\sigma_{\rm int}$), accounting for the remaining eleven parameters.

From the {\it Gaia} absolute magnitude versus color diagram in Sect.~\ref{sec:stars}, we selected $123$ high-quality white dwarfs located at $d < 100$ pc.  We restricted the analysis to distances closer than $100$ pc, where the interstellar extinction can be neglected \citep[e.g.][]{lucke78,lallement03,zucker22} and the observed J-PLUS magnitudes can be used therefore without correction from dust reddening. First, a simplified version of the model was run with an extra component to identify outliers, this is, white dwarfs that are far from the theoretical locus. This was performed in sequence, starting from the $z$ band and moving to shorter wavelengths. In each color-color diagram, the outliers were identified and excluded. From the initial sample of $123$ white dwarfs, we identified $14$ outliers. Second, the final joint Bayesian analysis of the locus for $109$ white dwarfs in the eleven color-color diagrams was performed to compute the final offsets $\Delta \mathcal{X}_{\rm WD}$. The result for the $u$ passband is presented in Fig.~\ref{fig:wdlocus_1}. The color-color diagrams for the other passbands are gathered in Appendix~\ref{app:wdlocus}.

\begin{figure}[t]
\centering
\resizebox{\hsize}{!}{\includegraphics{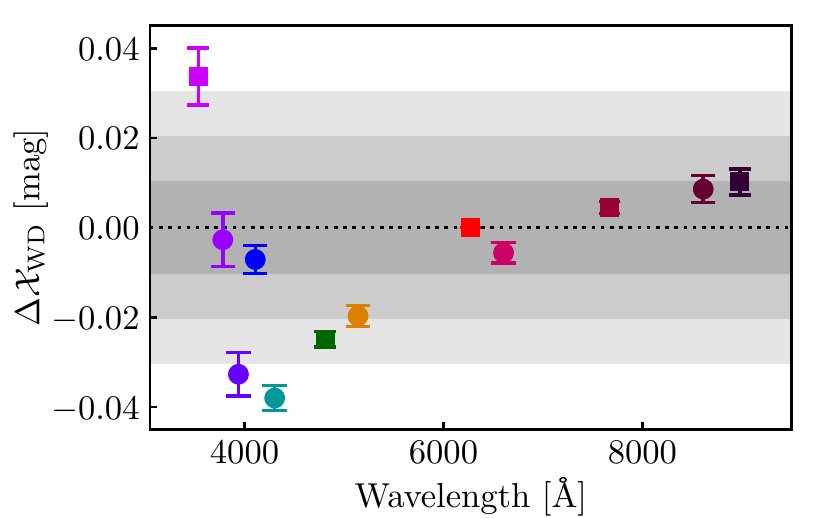}}
\caption{Zero point offset from the white dwarf locus ($\Delta \mathcal{X}_{\rm WD}$) for the J-PLUS filter system. The progressively lighter gray areas show differences of $0.01$, $0.02$, and $0.03$ mag, respectively.}
\label{fig:gaia_wd}
\end{figure}

The estimated parameters shared by all the color-color diagrams were $\mu = -0.237 \pm 0.010$, $s = 0.309 \pm 0.009$, $f_{\rm H} = 0.778 \pm 0.020$, and $\log {\rm g} = 8.08 \pm 0.03$. The obtained offsets from the white dwarf locus technique are summarized in Table~\ref{tab:wd_model} and Fig.~\ref{fig:gaia_wd}. We found a broad agreement with the {\it Gaia} BP/RP spectra scale, with differences below $0.04$ mag. There is a trend, with the offsets changing from $\Delta u_{\rm WD} = 0.034$ mag to $\Delta J0430_{\rm WD} = -0.038$ mag, then increasing again to $\Delta z_{\rm WD} = 0.010$ mag. We note that these are the residual differences after accounting for the magnitude and color terms presented in Sect.~\ref{sec:bprp}, and refer therefore to the median color of the calibration stars, $\bprp = 0.55$ mag. Thanks to the white dwarf locus, the $\mathcal{X}_{\rm G}$ magnitudes have been placed to the AB scale.

\begin{figure}[t]
\centering
\resizebox{\hsize}{!}{\includegraphics{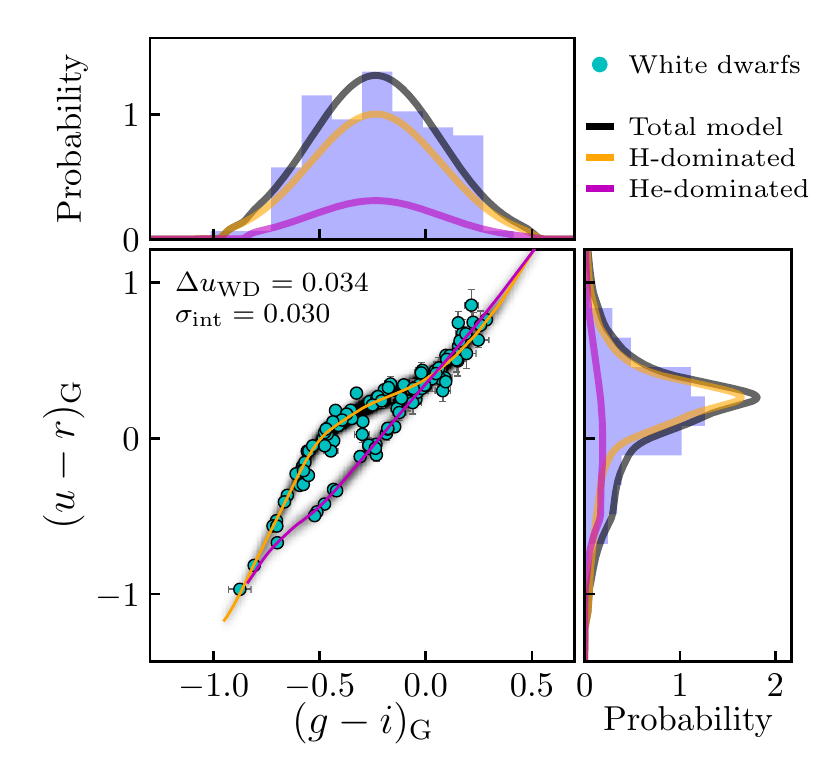}}
\caption{Color-color diagram $(u-r)_{\rm G}$ versus $(g-i)_{\rm G}$ of the $109$ high-quality white dwarfs at distance $d < 100$ pc in J-PLUS DR3. The solid lines show the theoretical loci for H- (orange) and He-dominated atmospheres (magenta). The gray scale shows the most probable model that describes the observations. The blue probability distributions above and to the right show the $(g-i)_{\rm G}$ and $(u - r)_{\rm G}$ projections of the data, respectively. The projections of the total, H-dominated, and He-dominated models are represented by the black, orange, and magenta lines. The values of the filter-dependent parameters $\sigma_{\rm int}$ and $\Delta \mathcal{X}_{\rm WD}$ are indicated in the panel.}
\label{fig:wdlocus_1}
\end{figure}

\subsection{Step 4: Absolute flux scale with PS1}\label{sec:deltar}
The white dwarf locus technique is able to provide the absolute color scale of the J-PLUS passbands, with the $r$ band used as reference. Because of the magnitude and color terms between BP/RP spectra and J-PLUS, we used the PS1 magnitudes in $r$ to set the absolute flux scale for the J-PLUS photometry.

We cross-matched the main sequence calibration stars with the PS1 DR1 catalog using a $1.5^{\arcsec}$ radius\footnote{The complete J-PLUS versus PS1 catalog can be found in the table \texttt{jplus.xmatch\_panstarrs\_dr1} within the J-PLUS database.}. Those sources with more than one counterpart in the PS1 catalog or without a valid photometric measurement on $gri$ PS1 passbands were discarded. We used the PS1 PSF magnitudes as reference \citep{ps1phot}.

We compared the homogenized magnitudes $r_{\rm G}$ from J-PLUS with the transformed $r$-band magnitudes from PS1. The transformation term accounts for the difference between the J-PLUS and PS1 passbands:
\begin{eqnarray}
\mathcal{C}_{\rm PS1} &=& g_{\rm PS1} - i_{\rm PS1},\\
T_r^{\rm PS1} &=&   \ \ \ \ \ 4.9 - \ \ 3.2 \times \mathcal{C}_{\rm PS1} + \ \ 8.2 \times \mathcal{C}^2_{\rm PS1}\ {\rm [mmag]}.
\end{eqnarray}
This transformation is valid at $0.4 < \mathcal{C}_{\rm PS1} < 1.4$ and only sources within this color range were used in the comparison. The details about the estimation of this transformation term are presented in \citet{clsj19jcal}.

The median of the differences between the magnitudes was computed for each pointing, providing the offset between the {\it Gaia} and PS1 photometric scales. The distribution of the differences for the $1\,642$ pointings in J-PLUS DR3 follows a Gaussian with a median of $4.3$ mmag and a dispersion  of $3.4$ mmag. Hence, we set $\Delta r_{\rm PS1} = 4.3$ mmag in Eq.~(\ref{eq:zp}) and assumed an uncertainty of $5$ mmag in this absolute flux scale. The accuracy of the absolute scale is tested in Sect.~\ref{sect:gd153}.

\begin{table} 
\caption{Estimated offsets to obtain the AB color scale of the J-PLUS passbands.} 
\label{tab:wd_model}
\centering 
        \begin{tabular}{l c c}
        \hline\hline\rule{0pt}{3ex} 
        Passband $(\mathcal{X})$   &   $\Delta \mathcal{X}_{\rm WD}$    &   $\sigma_{\rm int}$ \\\rule{0pt}{2ex} 
                &   [mmag]                & [mmag]   \\
        \hline\rule{0pt}{2ex}
        $u$             &$\ \ 33.7 \pm 6.4$             &$30 \pm 4$          \\ 
        $J0378$         &$\   -2.7 \pm 6.0$             &$31 \pm 4$          \\ 
        $J0395$         &$   -32.7 \pm 4.9$             &$24 \pm 4$          \\ 
        $J0410$         &$\   -7.1 \pm 3.1$             & $9 \pm 5$          \\ 
        $J0430$         &$   -38.0 \pm 2.8$             &$12 \pm 4$          \\ 
        $g$             &$   -24.9 \pm 1.7$             & $2 \pm 2$          \\ 
        $J0515$         &$   -19.7 \pm 2.3$             &$10 \pm 2$          \\ 
        $r$             &$\cdots$                       &$\cdots$            \\ 
        $J0660$         &$\   -5.6 \pm 2.3$             &$11 \pm 3$          \\ 
        $i$             &$\ \  4.5 \pm 1.3$             & $2 \pm 2$          \\ 
        $J0861$         &$\ \  8.6 \pm 3.0$             & $9 \pm 5$          \\ 
        $z$             &$\   10.2 \pm 2.9$             &$10 \pm 4$          \\ 
        \hline
\end{tabular}
\end{table}

\begin{figure*}[t]
\centering
\resizebox{0.32\hsize}{!}{\includegraphics{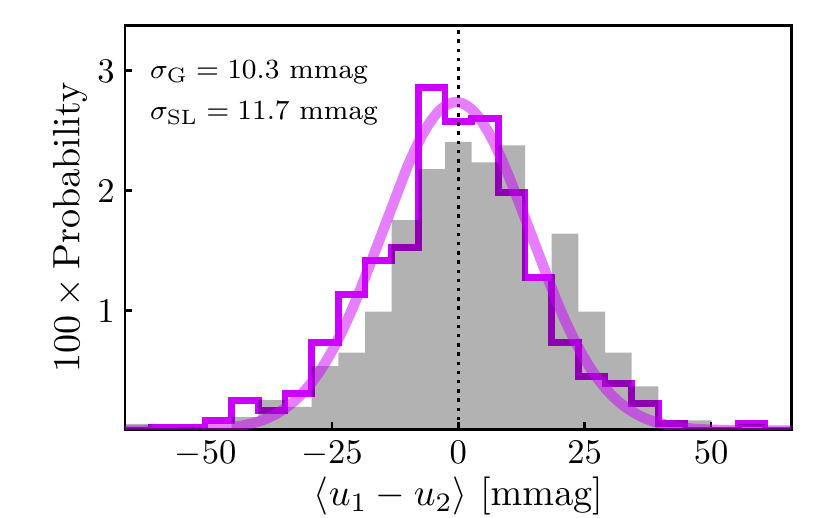}}
\resizebox{0.32\hsize}{!}{\includegraphics{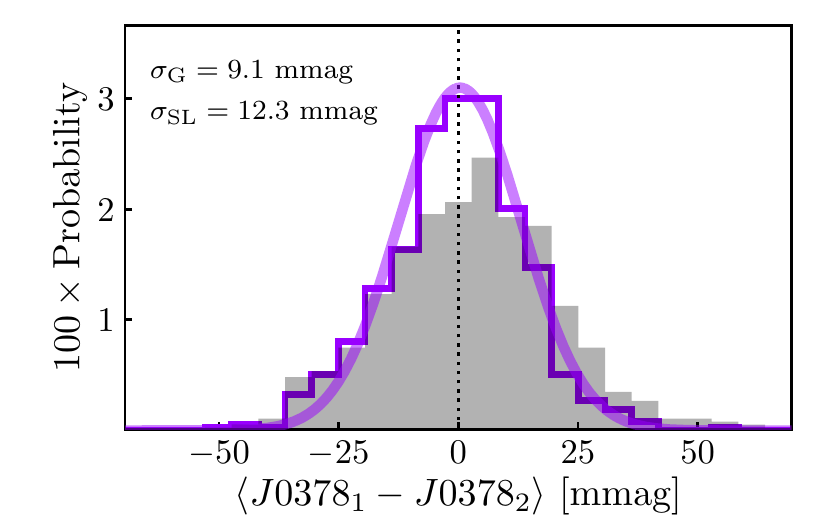}}
\resizebox{0.32\hsize}{!}{\includegraphics{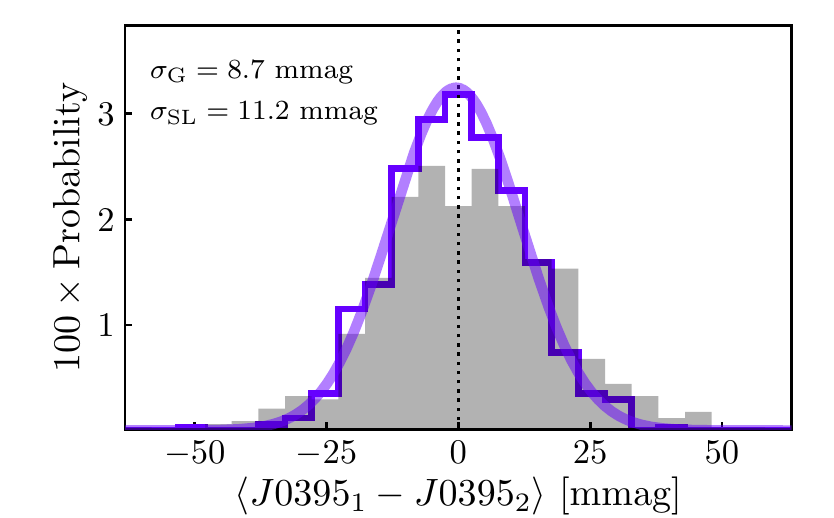}}

\resizebox{0.32\hsize}{!}{\includegraphics{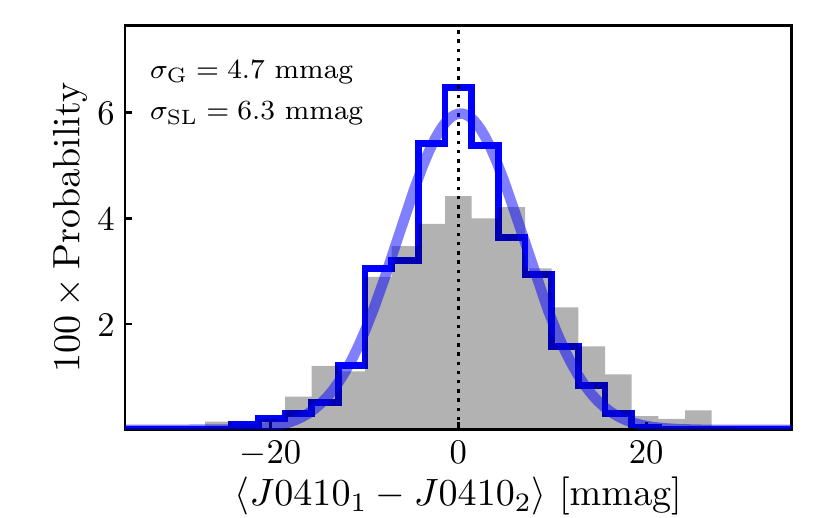}}
\resizebox{0.32\hsize}{!}{\includegraphics{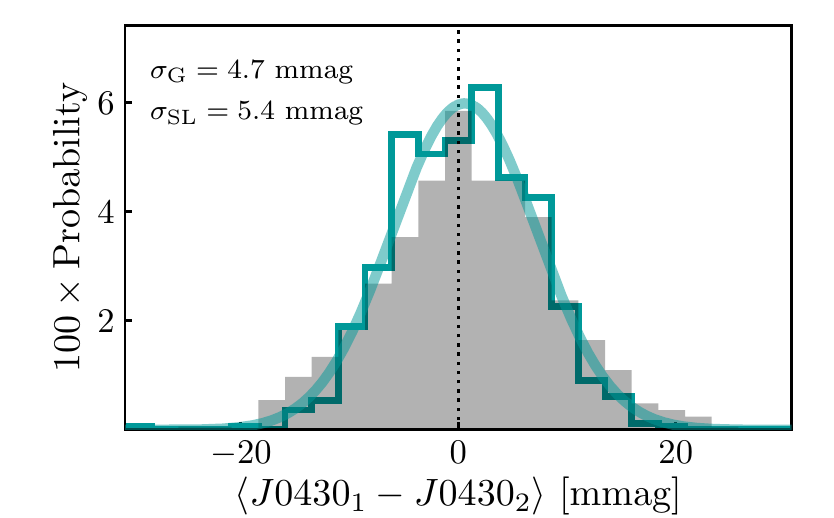}}
\resizebox{0.32\hsize}{!}{\includegraphics{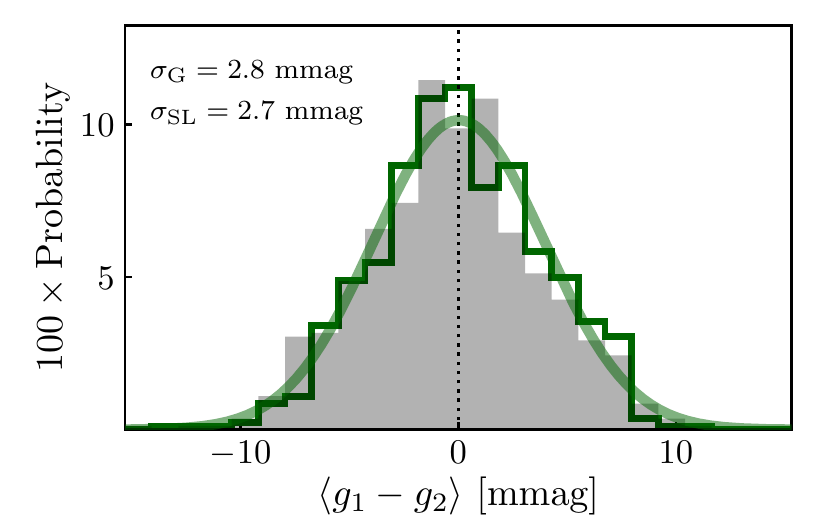}}

\resizebox{0.32\hsize}{!}{\includegraphics{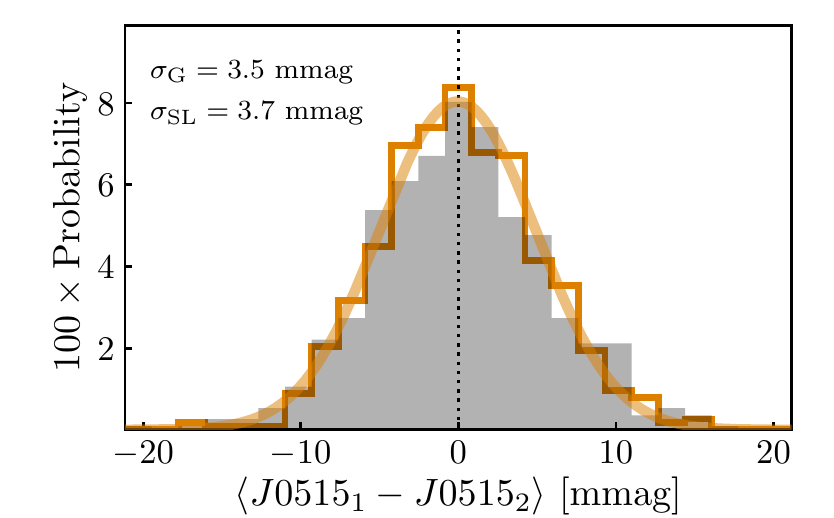}}
\resizebox{0.32\hsize}{!}{\includegraphics{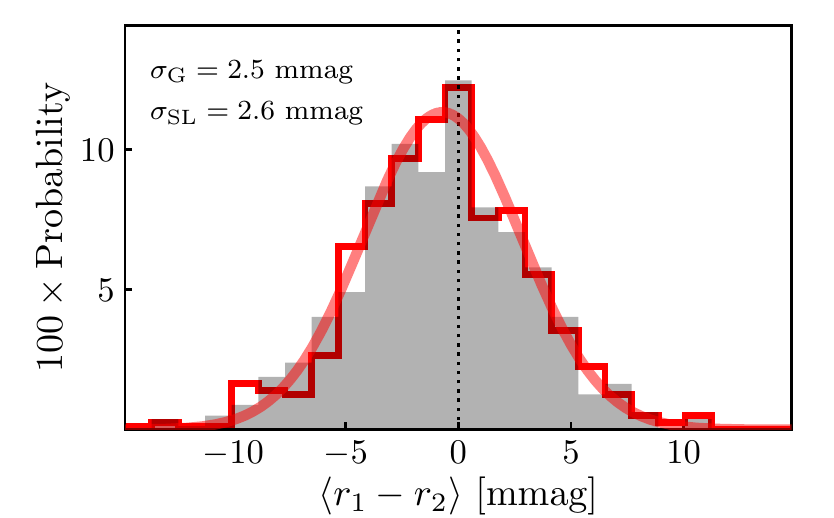}}
\resizebox{0.32\hsize}{!}{\includegraphics{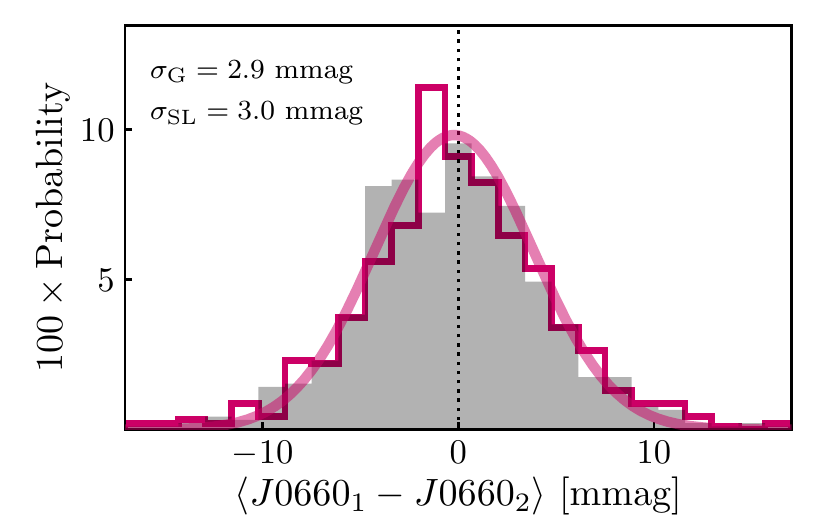}}

\resizebox{0.32\hsize}{!}{\includegraphics{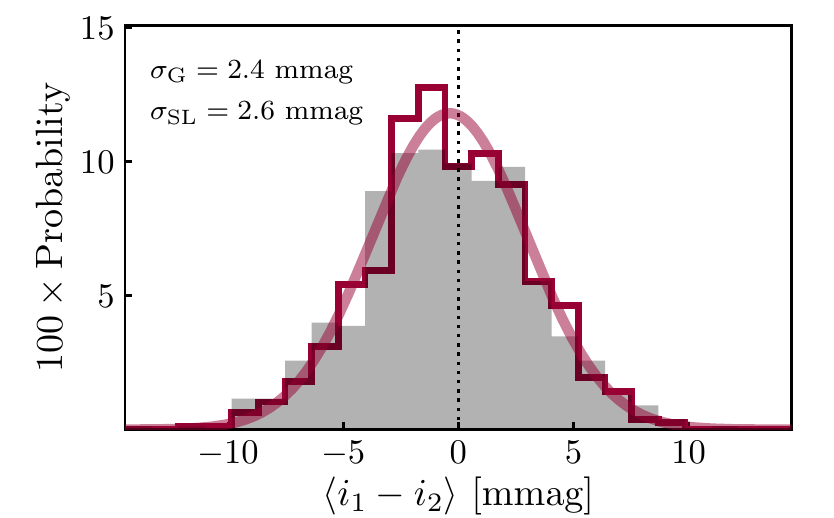}}
\resizebox{0.32\hsize}{!}{\includegraphics{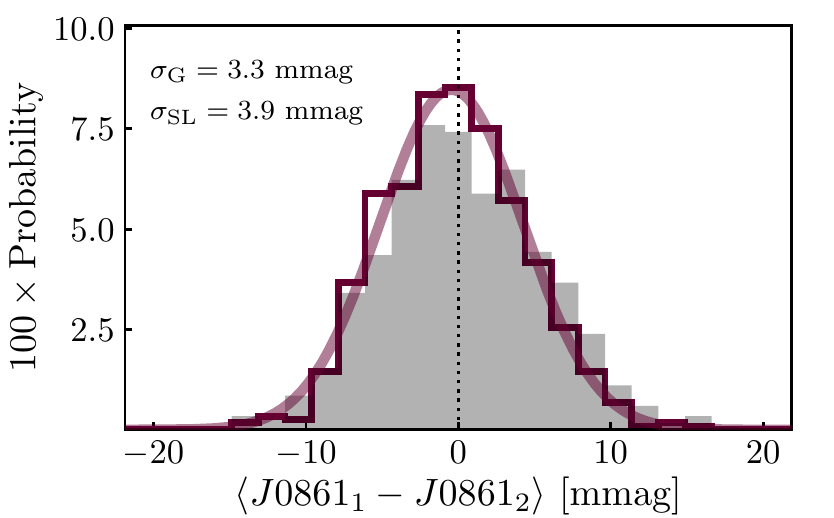}}
\resizebox{0.32\hsize}{!}{\includegraphics{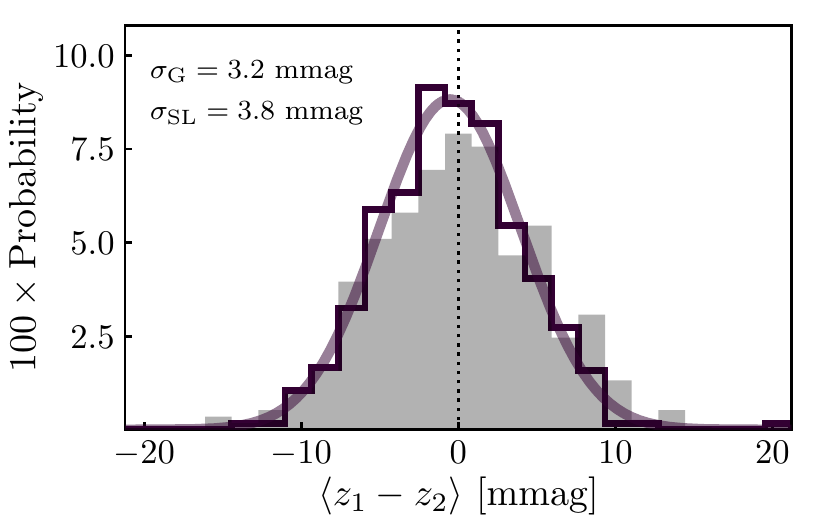}}
\caption{Distribution of median differences in the photometry of main sequence stars independently observed by two adjacent pointings with at least $25$ sources in common. The gray filled histogram shows the results obtained with the stellar locus regression technique, and the colored histogram using the {\it Gaia} BP/RP low-resolution spectra as reference. The gray and colored lines are the best Gaussian fits to the former and latest case, respectively. The precision in the calibration is labeled in the panels and was estimated as the dispersion of the fitted Gaussian divided by the square root of two. We present, from top to bottom and from left to right, the filters $u$, $J0378$, $J0395$, $J0410$, $J0430$, $g$, $J0515$, $r$, $J0660$, $i$, $J0861$, and $z$.
} 
\label{fig:difftile}
\end{figure*}

\section{Error budget}\label{sect:discussion}
This section is devoted to the error budget analysis in the J-PLUS DR3 calibration. We study the relative precision in the photometry in Sect.~\ref{sec:precision}, the relative accuracy across the surveyed area in Sect.~\ref{sec:accuracy}, and the absolute accuracy in Sect.~\ref{sect:gd153}.

\begin{figure*}[t]
\centering
\resizebox{0.32\hsize}{!}{\includegraphics{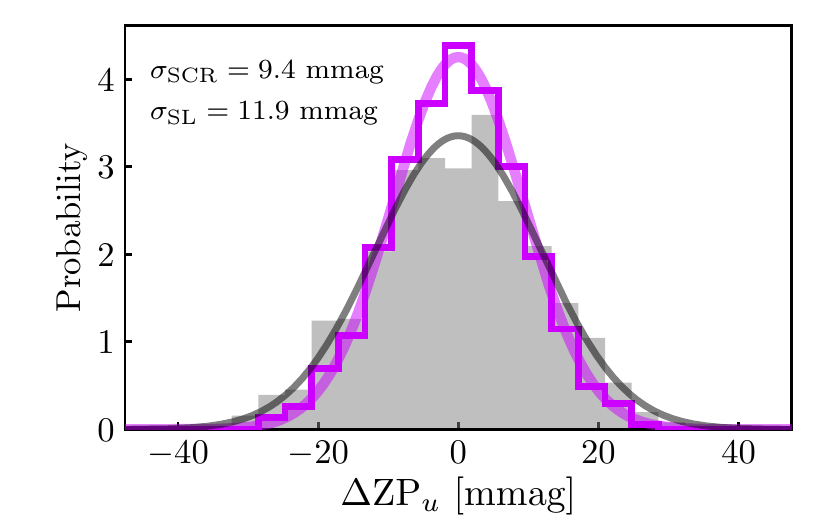}}
\resizebox{0.32\hsize}{!}{\includegraphics{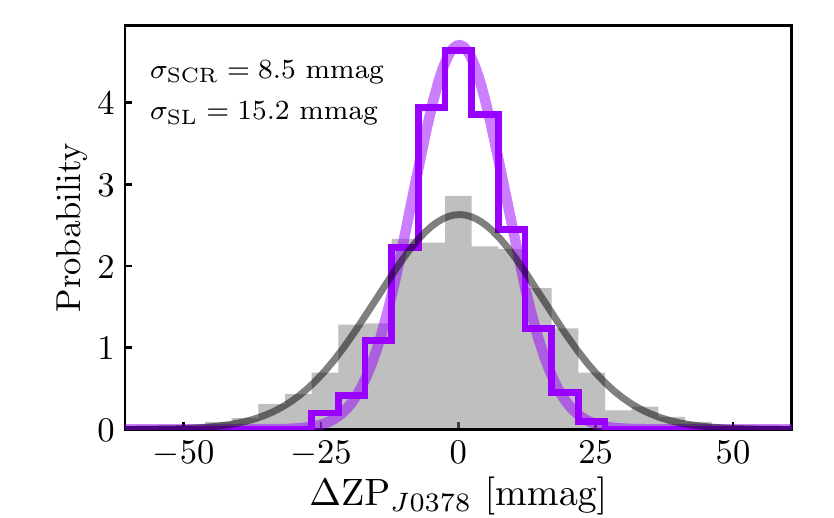}}
\resizebox{0.32\hsize}{!}{\includegraphics{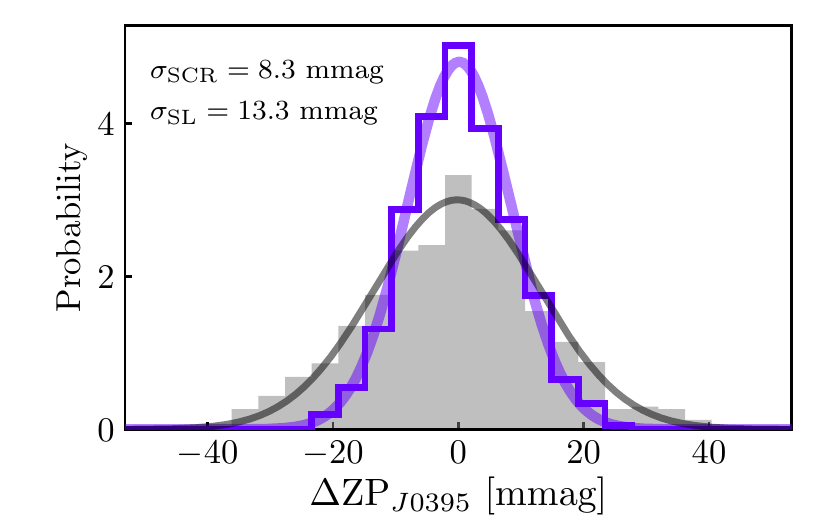}}

\resizebox{0.32\hsize}{!}{\includegraphics{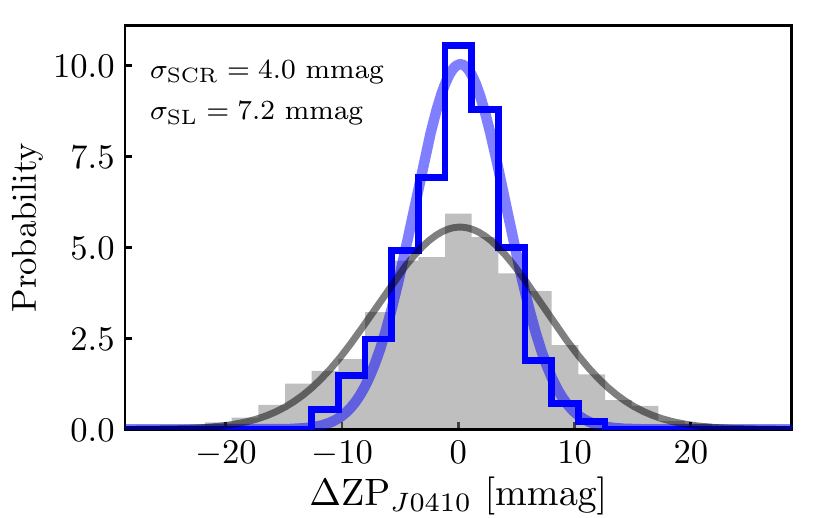}}
\resizebox{0.32\hsize}{!}{\includegraphics{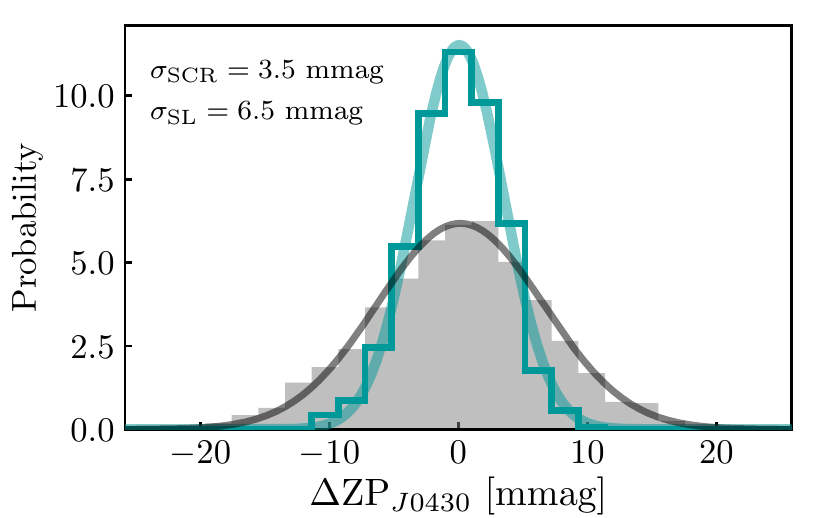}}
\resizebox{0.32\hsize}{!}{\includegraphics{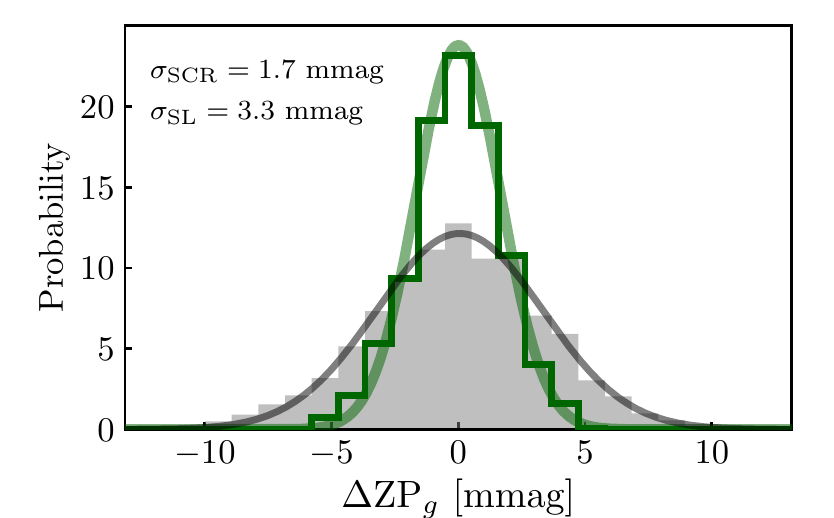}}

\resizebox{0.32\hsize}{!}{\includegraphics{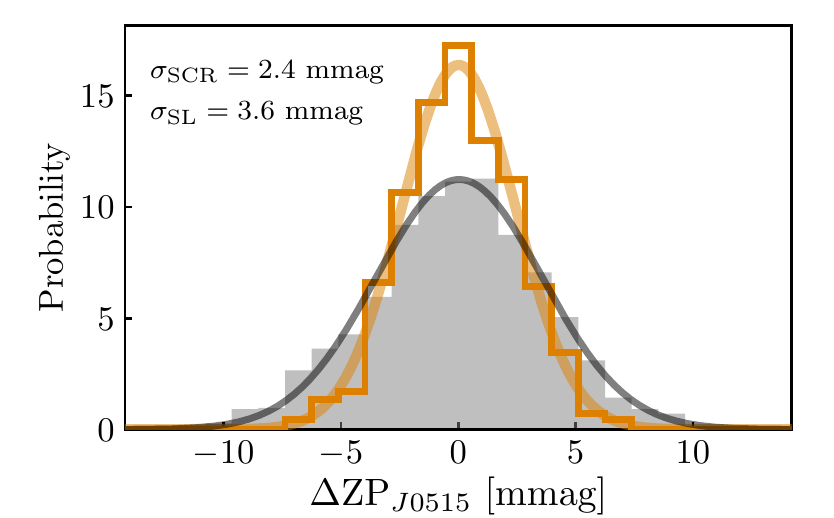}}
\resizebox{0.32\hsize}{!}{\includegraphics{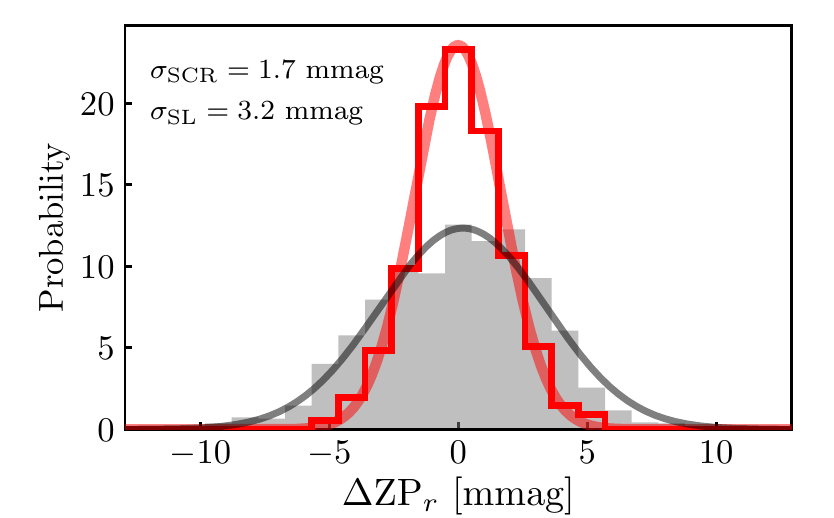}}
\resizebox{0.32\hsize}{!}{\includegraphics{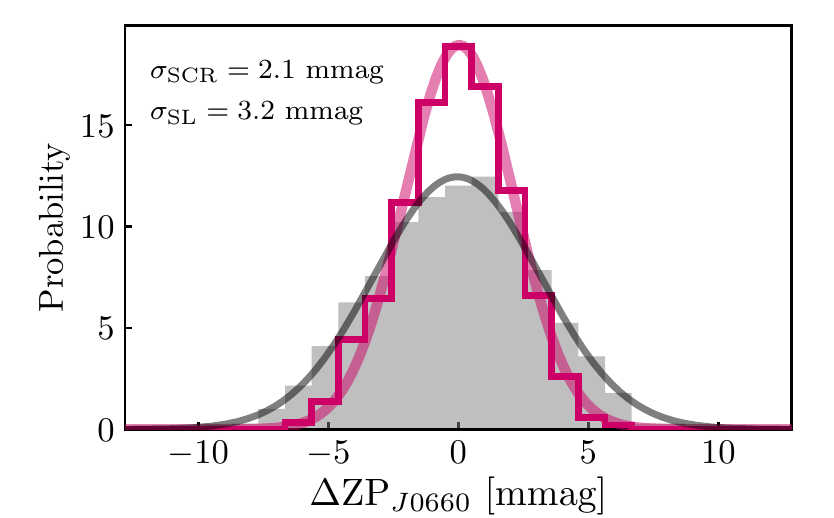}}

\resizebox{0.32\hsize}{!}{\includegraphics{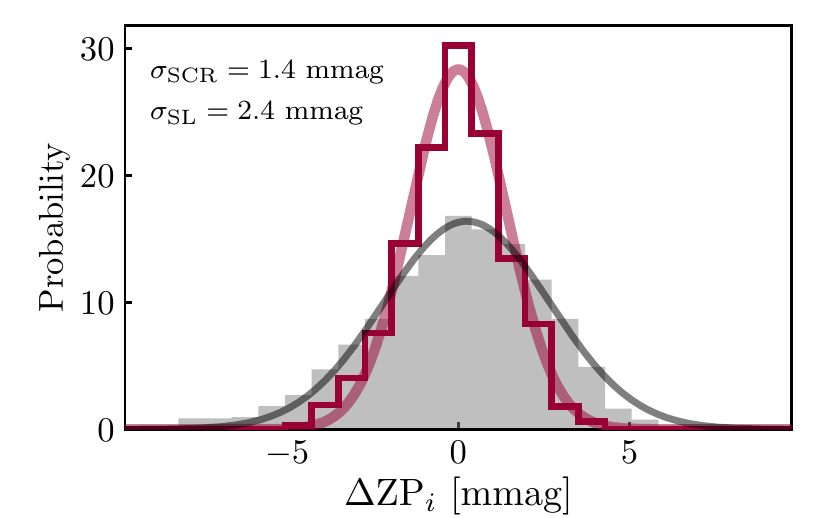}}
\resizebox{0.32\hsize}{!}{\includegraphics{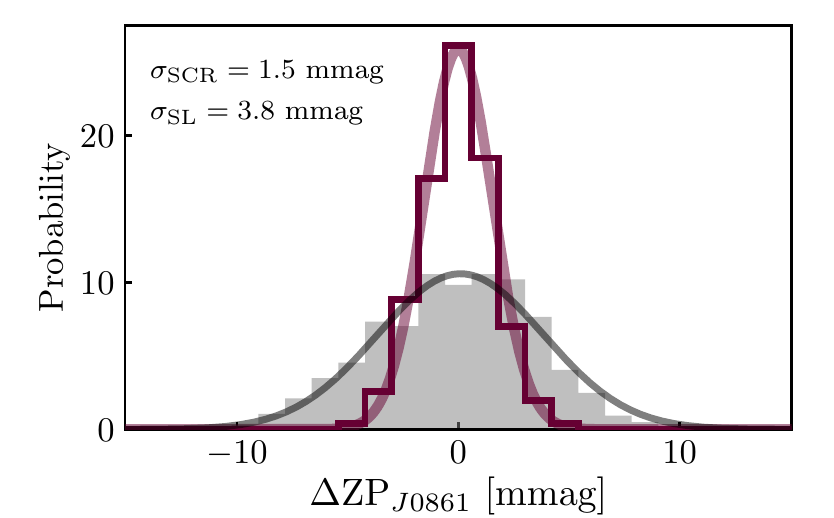}}
\resizebox{0.32\hsize}{!}{\includegraphics{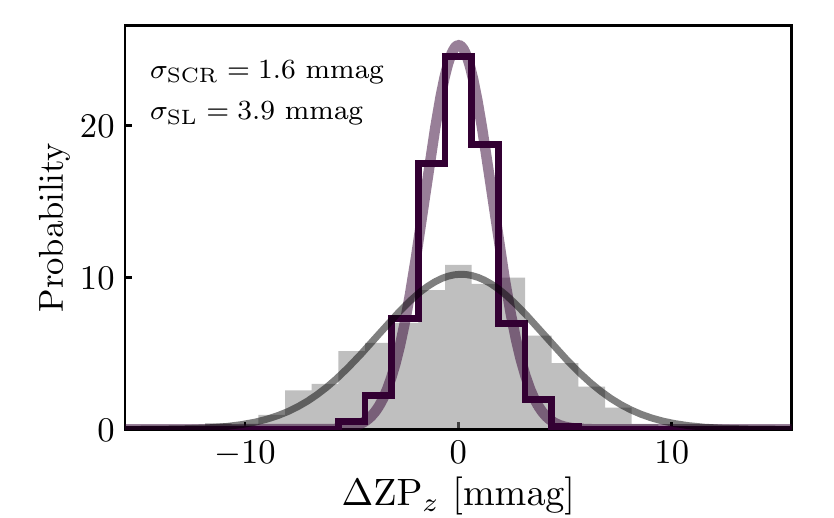}}
\caption{Distribution of the difference between the {\it Gaia}-based and the SCR-based zero points, $\Delta ZP_{\mathcal{X}}$. The gray filled histogram shows the comparison with the zero points obtained from the stellar locus regression technique applied to J-PLUS DR3, and the colored histogram using the {\it Gaia} BP/RP low-resolution spectra as reference. The gray and colored lines are the best Gaussian fits to the former and latest case, respectively. The accuracy in the calibration is labeled in the panels. We present, from top to bottom and from left to right, the filters $u$, $J0378$, $J0395$, $J0410$, $J0430$, $g$, $J0515$, $r$, $J0660$, $i$, $J0861$, and $z$.
} 
\label{fig:diffscr}
\end{figure*}

\subsection{Relative precision from overlapping areas}\label{sec:precision}
Adjacent J-PLUS pointings slightly overlap with each other. To measure the precision of the calibration, the photometry of calibration stars independently observed in two pointings was compared. The number of unique pointings pairs with overlap in J-PLUS DR3 is $4\,247$. For each pointing pair, we computed the difference between the two calibrated magnitudes of the common stars and estimated the median of the differences. To minimize the effect of the individual errors, only those pointing pairs with $25$ or more common sources were kept. This provided $670$ median differences. The targeted precision was obtained as $\sigma/\sqrt{2}$, where $\sigma$ was the measured dispersion of the differences distribution. The obtained precision is summarized in Fig.~\ref{fig:difftile} and Table~\ref{tab:jplus_calib}, being, on average, $\sim 9$ mmag in $u$, $J0378$, and $J0395$; $\sim 5$ mmag $J0410$ and $J0430$; and $\sim 3$ mmag in $g$, $J0515$, $r$, $J0660$, $i$, $J0861$, and $z$.

\begin{figure*}[t]
\centering
\resizebox{0.49\hsize}{!}{\includegraphics{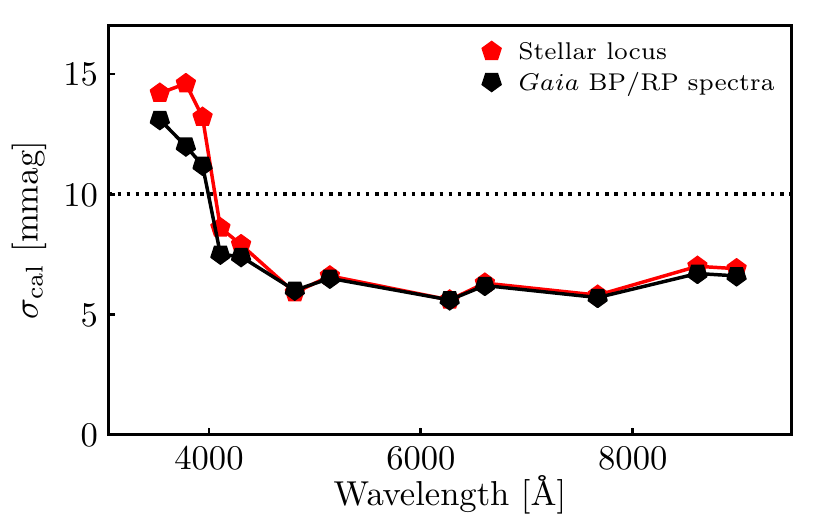}}
\resizebox{0.49\hsize}{!}{\includegraphics{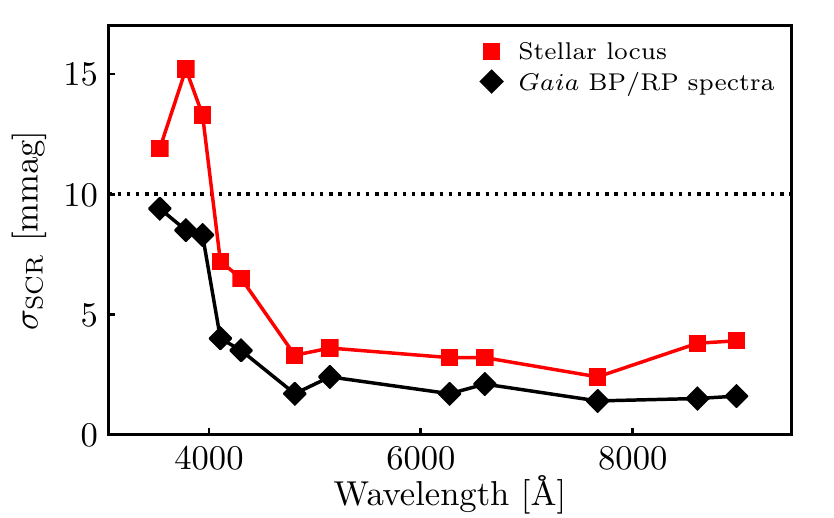}}
\caption{Summary of the relative precision ($\sigma_{\rm cal}$, {\it left panel}) and accuracy ($\sigma_{\rm SCR}$, {\it right panel}) estimated for the J-PLUS DR3 photometric calibration. Red and black symbols show the results obtained with the stellar locus technique and the BP/RP spectra from {\it Gaia}, respectively. The dotted line marks a 1\% level uncertainty.
} 
\label{fig:sigmazp}
\end{figure*}

We tested the change with respect to the stellar locus methodology, the reference calibration method in J-PLUS DR2 \citep{clsj19jcal,clsj21zsl}. Following the same definitions, we found an improvement of $\sim 20$\% in the precision at $\lambda < 4\,500$ \AA{}, and compatible results for the rest of the passbands (Figs.~\ref{fig:difftile} and \ref{fig:sigmazp}). We highlight the improvement found in the bluer passbands, where the signal of the BP/RP spectra is lower and a better performance than the stellar locus technique was not ensured. These results support the capabilities of the {\it Gaia} BP/RP spectra to obtain an homogeneous photometry across the sky and greatly simplifies the calibration procedure, with no previous information about neither the extinction nor the metallicity of the sources being required.

\begin{table*} 
\caption{Estimated error budget of the J-PLUS DR3 photometric calibration and final median zero points.}
\label{tab:jplus_calib}
\centering 
        \begin{tabular}{@{\extracolsep{4pt}}l c c c c c}
        \hline\hline\noalign{\smallskip}
           &  \multicolumn{3}{c}{Precision}  & Accuracy & \\
        \cline{2-4}\cline{5-5}\noalign{\smallskip}
        Passband    & $\sigma_{\rm G}$   & $\sigma_{\rm WD}$ & $\sigma_{\rm cal}$ & $\sigma_{\rm SCR}$&  $\langle {\rm ZP}_{\mathcal{X}}\rangle$\\
                &       [mmag]\tablefootmark{a}   &      [mmag]\tablefootmark{b}          &         [mmag]\tablefootmark{c}           & [mmag]\tablefootmark{d}   & [mag]\\

        \hline\noalign{\smallskip}
        $u$             &  10.3  &   6.4  & 13.1    &  9.4 & 21.10\\ 
        $J0378$         &   9.1  &   6.0  & 12.0    &  8.5 & 20.48\\ 
        $J0395$         &   8.7  &   4.9  & 11.2    &  8.3 & 20.36\\ 
        $J0410$         &   4.7  &   3.1  &  7.5    &  4.0 & 21.32\\ 
        $J0430$         &   4.7  &   2.8  &  7.4    &  3.5 & 21.38\\ 
        $g$             &   2.8  &   1.7  &  6.0    &  1.7 & 23.59\\ 
        $J0515$         &   3.5  &   2.3  &  6.5    &  2.4 & 21.56\\ 
        $r$             &   2.5  &$\cdots$&  5.6    &  1.7 & 23.64\\ 
        $J0660$         &   2.9  &   2.3  &  6.2    &  2.1 & 21.10\\ 
        $i$             &   2.4  &   1.3  &  5.7    &  1.4 & 23.34\\ 
        $J0861$         &   3.3  &   3.0  &  6.7    &  1.5 & 21.64\\ 
        $z$             &   3.2  &   2.9  &  6.6    &  1.6 & 22.78\\ 
        \hline 
\end{tabular}
\tablefoot{
\tablefoottext{a} {{\it Gaia} BP/RP low-resolution spectra and the plane correction to account for 2D variations along the CCD were used to homogenize the photometry. Precision estimated from duplicated main sequence stars in overlapping pointings (Sect.~\ref{sec:precision})}\\
\tablefoottext{b} {Uncertainty in the absolute color calibration from the Bayesian analysis of the white dwarf locus (Sect.~\ref{sec:wdlocus}).}\\
\tablefoottext{c} {Final precision in the J-PLUS DR3 flux calibration, $\sigma^2_{\rm cal} = \sigma^2_{\rm G} + \sigma^2_{\rm WD} + \sigma_r^2$, where $\sigma_r = 5$ mmag (Sect.~\ref{sec:deltar})}.\\
\tablefoottext{d}{Accuracy estimated from the comparison of the final calibration with results from the stellar color regression method (Sect.~\ref{sec:accuracy}).}
}
\end{table*}

\subsection{Relative accuracy along the surveyed area with the SCR method}\label{sec:accuracy}
The comparison of the photometry in adjacent pointings is not able to provide a measurement of the accuracy of the calibration along the surveyed area. The SCR method \citep{scr,huang21} uses the effective temperature, surface gravity, and metallicity from spectroscopy to match stars of the same properties (i.e. intrinsic colors) and ascribes the observed color differences to the effect of interstellar extinction. This permits the homogenization of the photometric solution by naturally accounting for temperature, gravity, metallicity, and extinction effects. The SCR has been used to validate and improve the photometric calibration of the Sloan Digital Sky Survey \citep{scr}, Pan-STARRS \citep{xiao22}, {\it Gaia} \citep{niu21_dr3, niu21_dr2}, or the Sky Mapper Southern Survey \citep{huang21}; reaching an accuracy better than 1\%. 

Using the atmospheric parameters from the Large Sky Area Multi-Object Fiber Spectroscopic Telescope (LAMOST, \citealt{lamost}) DR7\footnote{\url{http://www.lamost.org/dr7}}, the SCR method was applied to J-PLUS DR3. Due to the large sky coverage of LAMOST DR7, $1\,481$ (90\%) pointings were calibrated. A detailed application and analysis of the SCR calibration is beyond the scope of the present paper and will be presented in a forthcoming work.

We found that the difference between the {\it Gaia}-based and the SCR-based zero points follow a Gaussian distribution with dispersion $\sigma_{\rm SCR}$, as reported in Table~\ref{tab:jplus_calib}. The dispersion is $\sim 9$ mmag in $u$, $J0378$, and $J0395$; $\sim 4$ mmag in $J0410$ and $J0430$; and $\sim 2$ mmag in the rest of the J-PLUS passbands (Fig.~\ref{fig:diffscr}). The origin of this dispersion is related to the treatment of the interstellar extinction in SCR, the limitations in the all-sky homogeneity of the BP/RP spectra, and the inherent statistical dispersion of each method.

As in the previous section, the accuracy using BP/RP spectra improves with respect to the use of the stellar locus technique. The dispersion between the zero points based on the stellar locus and the SCR is systematically higher (Figs.~\ref{fig:diffscr} and \ref{fig:sigmazp}). There is a general improvement of $\sim 40$\% in the accuracy when the BP/RP spectra were used as reference. Again, these results confirm that the {\it Gaia} low-resolution spectra are a competitive choice to perform the homogenization of large area, multi-filter surveys with minimum assumptions even at $\lambda < 4\,500$ \AA{}.

We note that the accuracy and the precision of the calibration present comparable figures for each passband. This suggests that the current methodology may be close to pure random uncertainties and that residual systematic differences should be below $1$\%. The comparison with the independent SCR method provides a proxy for the accuracy in the J-PLUS photometry, which we set at a percentage level or better for all the J-PLUS passbands.

\subsection{Absolute precision in the flux calibration}\label{sect:gd153}
\begin{figure*}[t]
\centering
\resizebox{0.49\hsize}{!}{\includegraphics{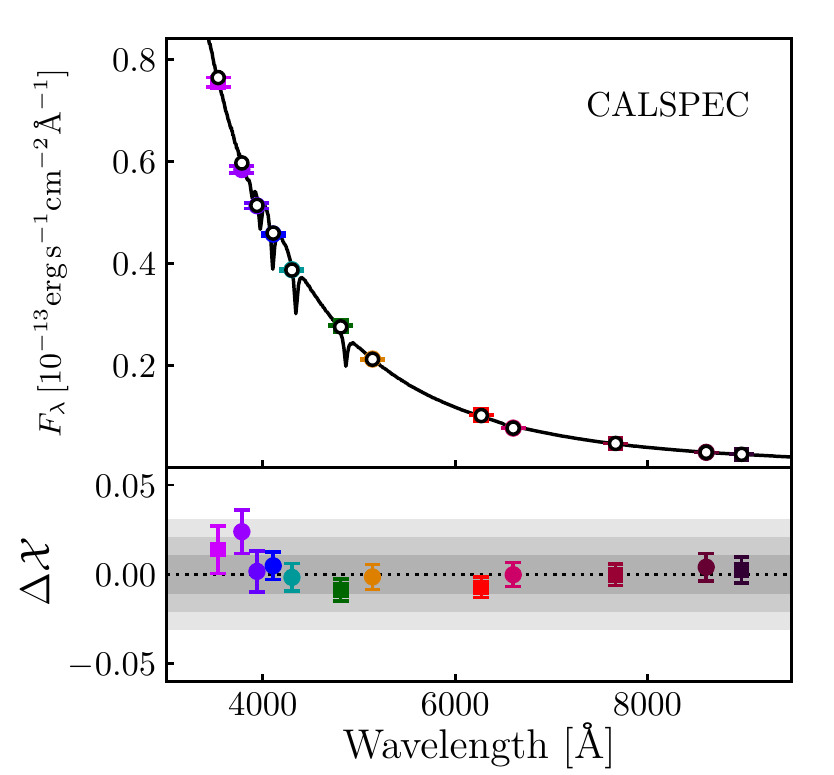}}
\resizebox{0.49\hsize}{!}{\includegraphics{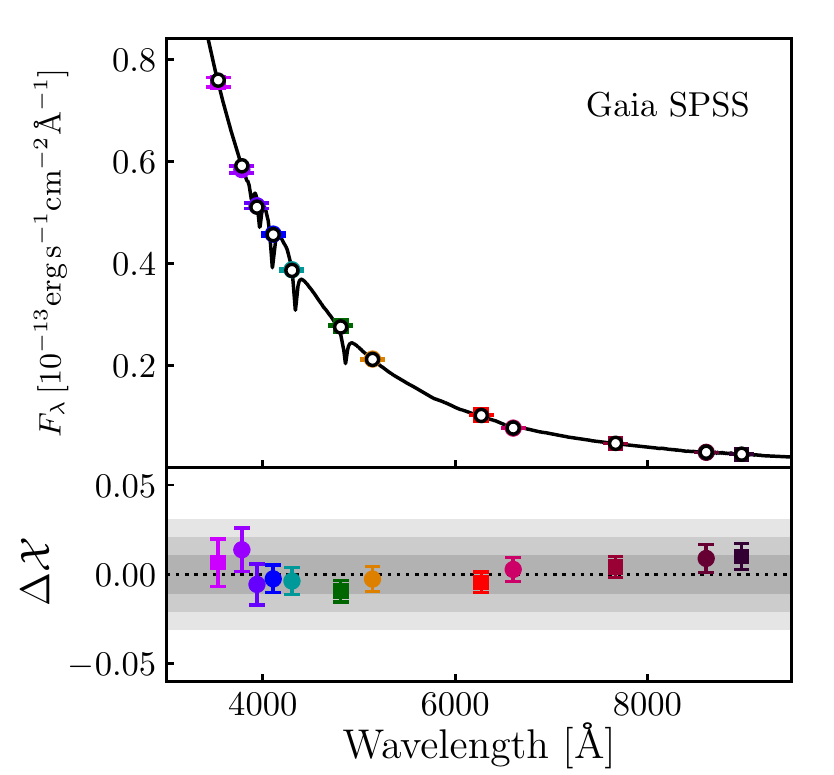}}
\caption{Comparison between the J-PLUS DR3 photometry ($\mathcal{X}_{\rm J-PLUS}$, colored points) of GD 153 and its synthetic photometry ($\mathcal{X}_{\rm standard}$, white dots) estimated from the standard spectra in CALSPEC ({\it left panel}) and {\it Gaia} SPSS ({\it right panel}). In both panels, the standard spectrum is shown with the black solid line. The magnitude difference $\Delta \mathcal{X} = \mathcal{X}_{\rm J-PLUS} - \mathcal{X}_{\rm standard}$ is shown in the lower panels. The dotted line marks a zero difference. The progressively lighter gray areas show differences of 0.01, 0.02, and 0.03 mag, respectively.}
\label{fig:sss_gd153}
\end{figure*}

We tested the absolute flux calibration in J-PLUS DR3 by comparing the final photometry with the synthetic photometry of the spectroscopic standard star GD $153$. This white dwarf is one of the three calibration pillars from the Hubble Space Telescope (HST) and it was observed as part of J-PLUS DR3. The $r-$band magnitude of GD $153$ in J-PLUS is $r = 13.59$ mag, so its photometry is dominated by calibration uncertainties with small photon counting errors. We found additional spectro-photometric standard stars observed by J-PLUS DR3. The individual results from these sources are noisier than for GD $153$, with similar average results. Moreover, the spectra of these extra standards are calibrated using the three HST pillars as reference.

We used the GD $153$ reference spectra from CALSPEC\footnote{\url{https://archive.stsci.edu/hlsps/reference-atlases/cdbs/current_calspec/gd153_stiswfcnic_003.fits}} \citep{calspec,bohlin20} and from the {\it Gaia} spectro-photometric standard stars (SPSS) survey\footnote{\url{http://gaiaextra.ssdc.asi.it:8900/reduced/2/SPSSpublic/V2.SPSS003.ascii}} \citep{gaia_spss_i,gaia_spss_v}. The results are presented in Fig.~\ref{fig:sss_gd153}. We found a remarkable 1\% agreement in all the passbands between the reference spectra and the J-PLUS photometry.

The situation has improved with respect to J-PLUS DR2, where the same test was performed \citep{clsj21zsl}. A difference of 3\% for the $u$ band, of 2\% for $J0378$, $J0395$, $J0410$, and $J0430$; and below 1\% for $g$, $J0515$, $r$, $J0660$, $i$, $J0861$, and $z$ was found. The new homogenization process based on {\it Gaia} BP/RP spectra and the use of dust-free white dwarfs in the estimation of the color scale have significantly decreased the discrepancies in the passbands at $\lambda < 4\,500$ \AA{}, pushing the absolute accuracy to the one per cent level in the complete J-PLUS filter system.

\section{Summary and conclusions}\label{sec:summary}
We presented the photometric calibration of the J-PLUS DR3 twelve optical passbands across $3\,284$ deg$^2$ of the northern sky. Synthetic photometry derived from the BP/RP spectra released as part of the {\it Gaia} DR3 for nearly $1.5$ million main sequence stars was used to homogenize the photometric solution. The AB color scale was derived using the locus of $109$ white dwarfs closer than $100$ pc, for which interstellar extinction can be neglected. Finally, the absolute flux scale was anchored to the Pan-STARRS photometry in the $r$ band. 

The relative precision in the calibration, measured from repeated sources in the overlapping areas between pointings and including absolute color and flux scale uncertainties, is $\sim 12$ mmag in $u$, $J0378$, and $J0395$; and $\sim 7$ mmag in $J0410$, $J0430$, $g$, $J0515$, $r$, $J0660$, $i$, $J0861$, and $z$. We found a $\sim 20\%$ improvement with respect to the stellar locus technique applied in DR2 for passbands with $\lambda < 4\,500$ \AA{}.

The relative accuracy was estimated by comparison with the stellar color regression methodology. We found a dispersion between both methods of $\sim 9$ mmag in $u$, $J0378$, and $J0395$; $\sim 4$ mmag in $J0410$ and $J0430$; and $\sim 2$ mmag in $g$, $J0515$, $r$, $J0660$, $i$, $J0861$, and $z$. There is a general $\sim 40$\% improvement when compared with the stellar locus technique. This demonstrates the capabilities of BP/RP spectra as a high-quality reference to homogenize ground-based optical photometry.

Finally, the absolute precision in the J-PLUS flux scale was set at $1$\% in all the passbands from the comparison with the spectro-photometric standard star GD $153$.

We conclude that the combination of the synthetic photometry derived from {\it Gaia} BP/RP spectra, used to homogenize the photometric solution, and the white dwarf locus, to retrieve the absolute color AB scale, is able to provide a photometric calibration for large area multi-filter optical surveys with one per cent (and below) accuracy and precision. The absolute calibration of the BP/RP spectra is expected to improve in the future {\it Gaia} data releases, to be tested with the decrease in the amplitude of the magnitude and color transformation functions with respect to the J-PLUS photometry and of the offsets requested by the white dwarf locus approach.

As a technical application, the all-sky coverage with BP/RP spectra would permit the photometric calibration of T80Cam images in quasi-real time, providing an estimation of the atmosphere's transparency to improve the queue execution of observing programs with different requirements.

\begin{acknowledgements}
We dedicate this paper to the memory of our six IAC colleagues and friends who met with a fatal accident in Piedra de los Cochinos, Tenerife, in February 2007, with  special thanks to Maurizio Panniello, whose teachings of \texttt{python} were so important for this paper.


We thank the relevant discussions with the J-PLUS collaboration members. 

Based on observations made with the JAST80 telescope at the Observatorio Astrof\'{\i}sico de Javalambre (OAJ), in Teruel, owned, managed, and operated by the Centro de Estudios de F\'{\i}sica del Cosmos de Arag\'on. We acknowledge the OAJ Data Processing and Archiving Unit (UPAD) for reducing the OAJ data used in this work.

Funding for the J-PLUS Project has been provided by the Governments of Spain and Arag\'on through the Fondo de Inversiones de Teruel; the Aragonese Government through the Research Groups E96, E103, E16\_17R, and E16\_20R; the Spanish Ministry of Science and Innovation (MCIN/AEI/10.13039/501100011033 y FEDER, Una manera de hacer Europa) with grants PID2021-124918NB-C41, PID2021-124918NB-C42, PID2021-124918NA-C43, and PID2021-124918NB-C44; the Spanish Ministry of Science, Innovation and Universities (MCIU/AEI/FEDER, UE) with grants PGC2018-097585-B-C21 and PGC2018-097585-B-C22; the Spanish Ministry of Economy and Competitiveness (MINECO) under AYA2015-66211-C2-1-P, AYA2015-66211-C2-2, AYA2012-30789, and ICTS-2009-14; and European FEDER funding (FCDD10-4E-867, FCDD13-4E-2685). The Brazilian agencies FINEP, FAPESP, and the National Observatory of Brazil have also contributed to this project.

J.~M.~C. was funded by the Spanish MICIN/AEI/10.13039/501100011033 and by "ERDF A way of making Europe" by the “European Union” through grants RTI2018-095076-B-C21 and PID2021-122842OB-C21, and the Institute of Cosmos Sciences University of Barcelona (ICCUB, Unidad de Excelencia ’Mar\'{\i}a de Maeztu’) through grant CEX2019-000918-M.

J.~V. acknowledges the technical members of the UPAD for their invaluable work: Juan Castillo, Tamara Civera, Javier Hern\'andez, \'Angel L\'opez, Alberto Moreno, and David Muniesa.

P.-E.~T. has received funding from the European Research Council under the European Union’s Horizon 2020 research and innovation programmes n. 677706 (WD3D) and n. 101002408 (MOS100PC).

P.~C. acknowledges support from Fundaç\~ao de Amparo \`a Pesquisa do Estado de S\~ao Paulo (FAPESP 21/08813-7) and  Conselho Nacional de Desenvolvimento Cient\'{\i}fico e Tecnol\'ogico (CNPq 310555/2021-3).

A.~E., A.~d.~P., H.~D.~S., and J.~A.~F.~O. acknowledge the financial support from the Spanish Ministry of Science and Innovation and the European Union - NextGenerationEU through the Recovery and Resilience Facility project ICTS-MRR-2021-03-CEFCA.

F.~J.~E. acknowledges financial support by ESA (SoW SCI-OO-SOW-00371).

This work has made use of data from the European Space Agency (ESA) mission {\it Gaia} (\url{https://www.cosmos.esa.int/gaia}), processed by the {\it Gaia} Data Processing and Analysis Consortium (DPAC, \url{https://www.cosmos.esa.int/web/gaia/dpac/consortium}). Funding for the DPAC has been provided by national institutions, in particular the institutions participating in the {\it Gaia} Multilateral Agreement. This job has made use of the \texttt{Python} package \texttt{GaiaXPy}, developed and maintained by members of the Gaia Data Processing and Analysis Consortium (DPAC), and in particular, Coordination Unit $5$ (CU5), and the Data Processing Centre located at the Institute of Astronomy, Cambridge, UK (DPCI).

The Pan-STARRS1 Surveys (PS1) and the PS1 public science archive have been made possible through contributions by the Institute for Astronomy, the University of Hawaii, the Pan-STARRS Project Office, the Max-Planck Society and its participating institutes, the Max Planck Institute for Astronomy, Heidelberg, and the Max Planck Institute for Extraterrestrial Physics, Garching, The Johns Hopkins University, Durham University, the University of Edinburgh, the Queen's University Belfast, the Harvard-Smithsonian Center for Astrophysics, the Las Cumbres Observatory Global Telescope Network Incorporated, the National Central University of Taiwan, the Space Telescope Science Institute, the National Aeronautics and Space Administration under Grant No. NNX08AR22G issued through the Planetary Science Division of the NASA Science Mission Directorate, the National Science Foundation Grant No. AST-1238877, the University of Maryland, Eotvos Lorand University (ELTE), the Los Alamos National Laboratory, and the Gordon and Betty Moore Foundation.

Guoshoujing Telescope (the Large Sky Area Multi-Object Fiber Spectroscopic Telescope LAMOST) is a National Major Scientific Project built by the Chinese Academy of Sciences. Funding for the project has been provided by the National Development and Reform Commission. LAMOST is operated and managed by the National Astronomical Observatories, Chinese Academy of Sciences.

This research made use of \texttt{Astropy}, a community-developed core \texttt{Python} package for Astronomy \citep{astropy}, and \texttt{Matplotlib}, a 2D graphics package used for \texttt{Python} for publication-quality image generation across user interfaces and operating systems \citep{pylab}.
\end{acknowledgements}

\bibliographystyle{aa}
\bibliography{biblio}

\begin{appendix}

\section{Magnitude and color transformations}\label{app:tfilter}
\begin{figure*}[t]
\centering
\resizebox{0.245\hsize}{!}{\includegraphics{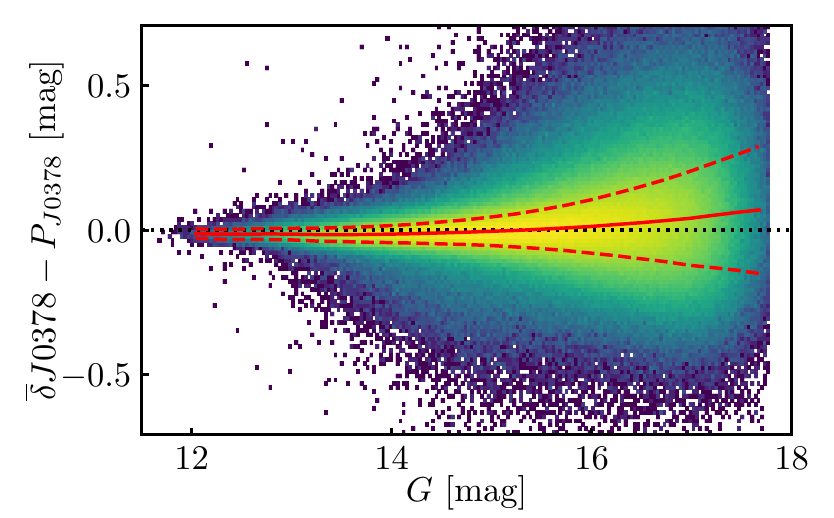}}
\resizebox{0.245\hsize}{!}{\includegraphics{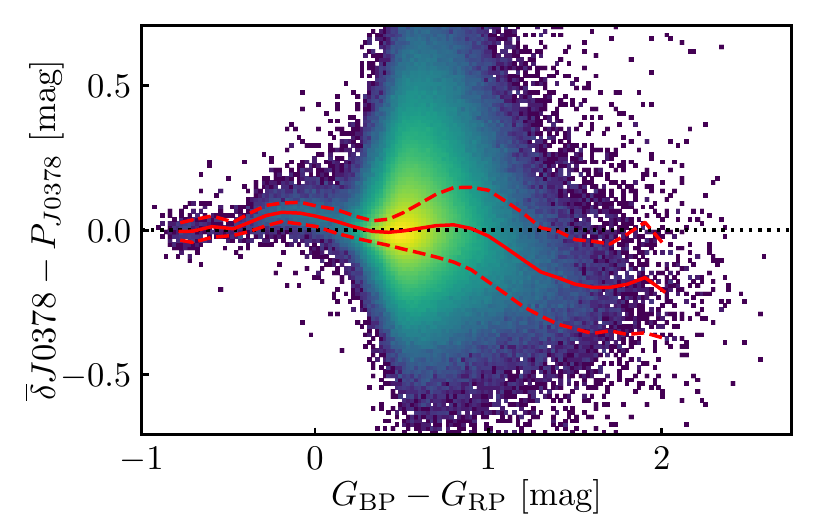}}
\resizebox{0.245\hsize}{!}{\includegraphics{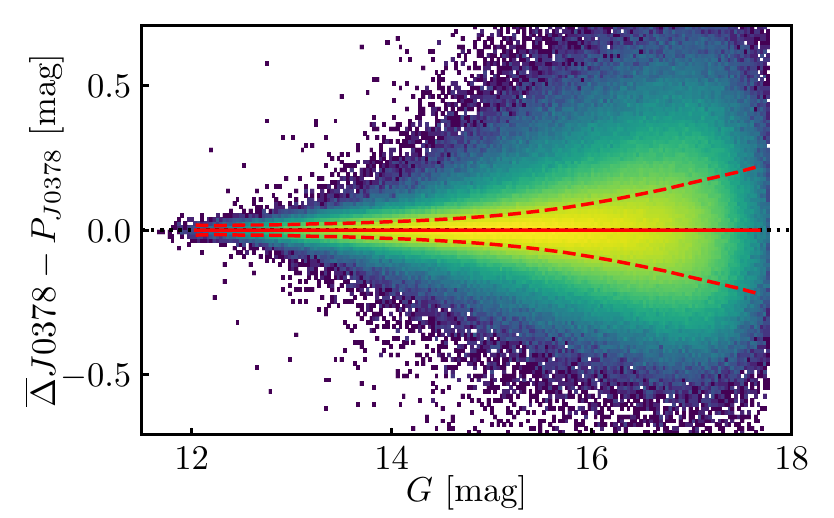}}
\resizebox{0.245\hsize}{!}{\includegraphics{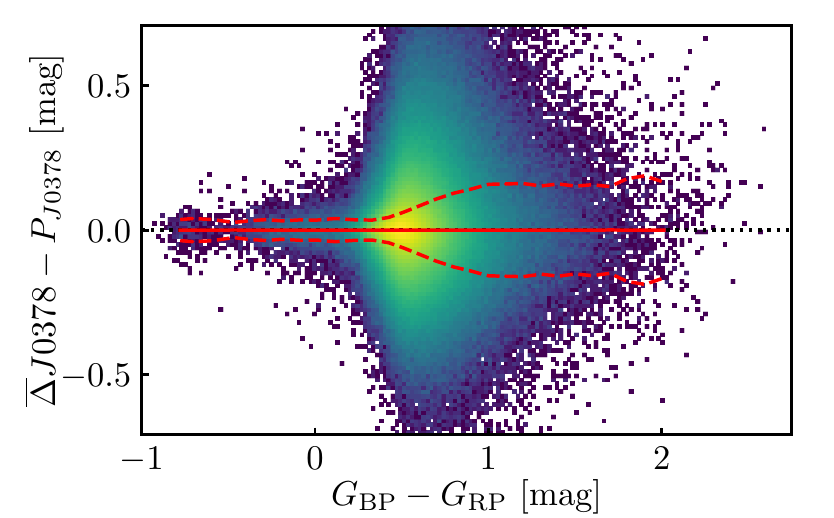}}\\
\resizebox{0.245\hsize}{!}{\includegraphics{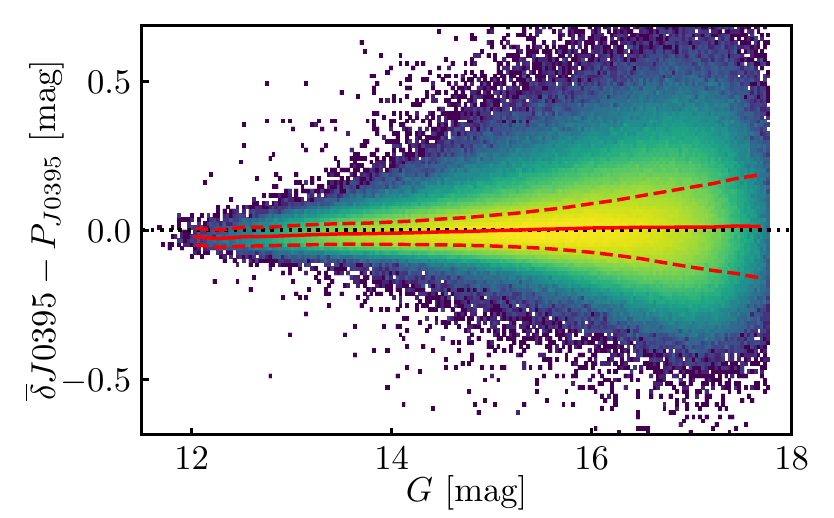}}
\resizebox{0.245\hsize}{!}{\includegraphics{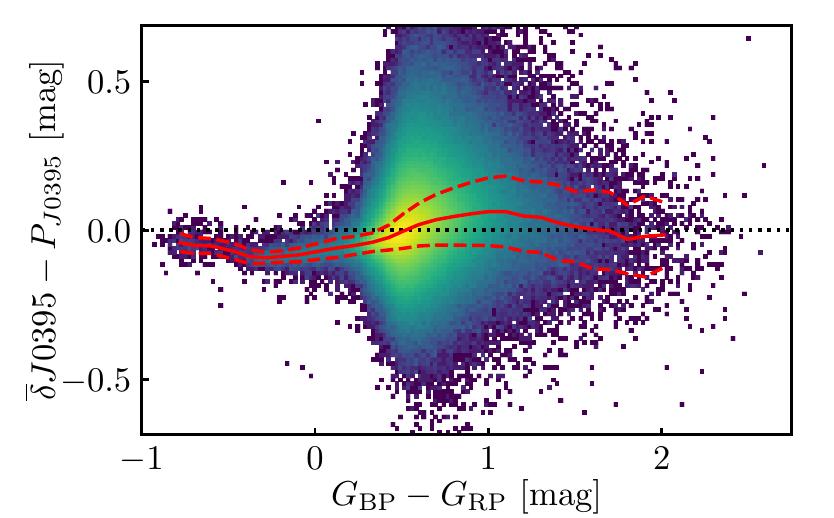}}
\resizebox{0.245\hsize}{!}{\includegraphics{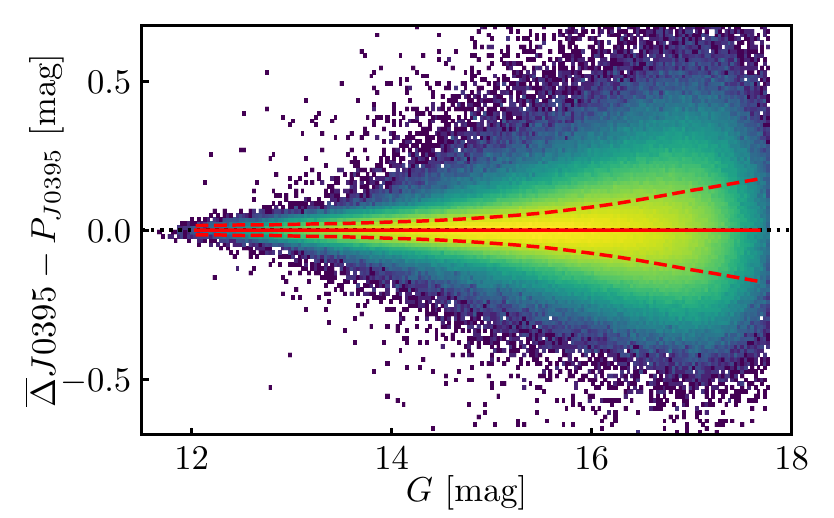}}
\resizebox{0.245\hsize}{!}{\includegraphics{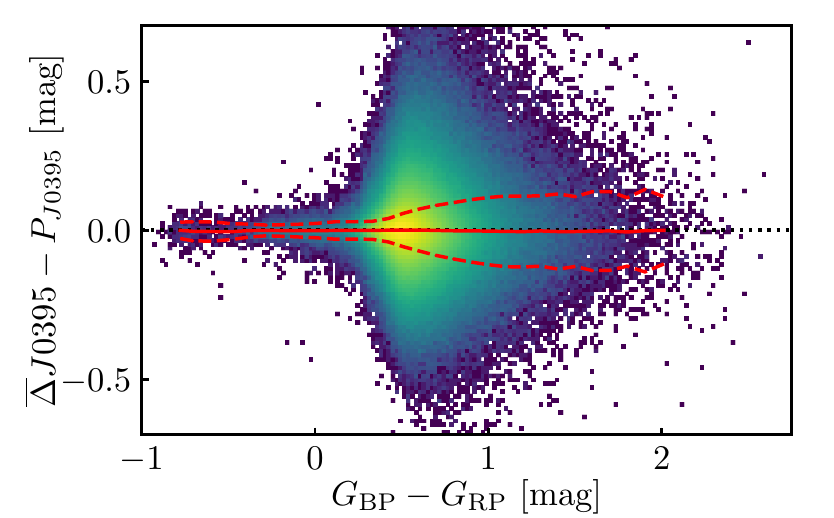}}\\
\resizebox{0.245\hsize}{!}{\includegraphics{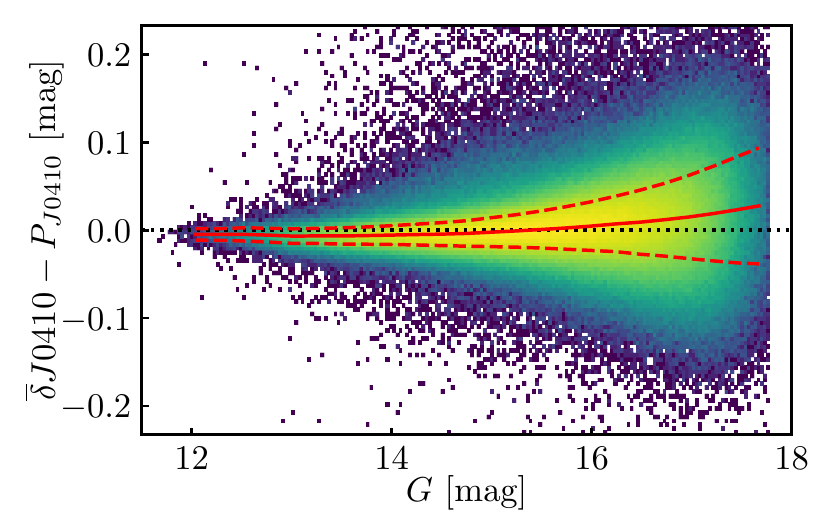}}
\resizebox{0.245\hsize}{!}{\includegraphics{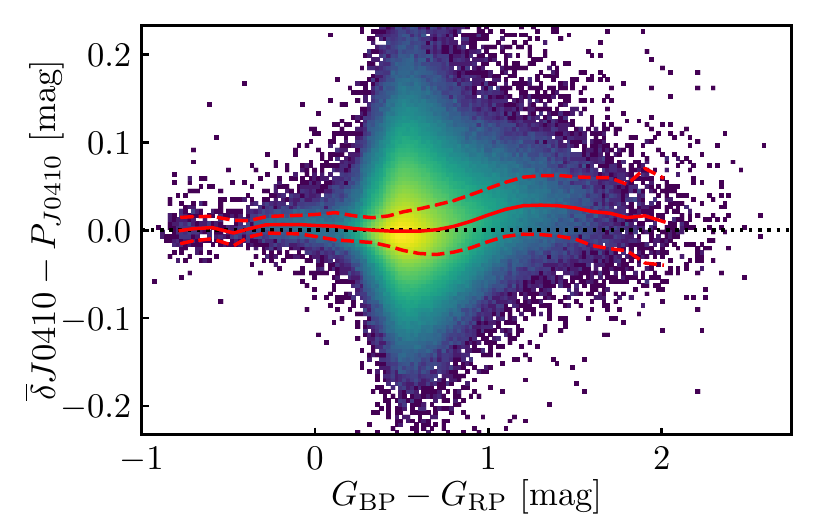}}
\resizebox{0.245\hsize}{!}{\includegraphics{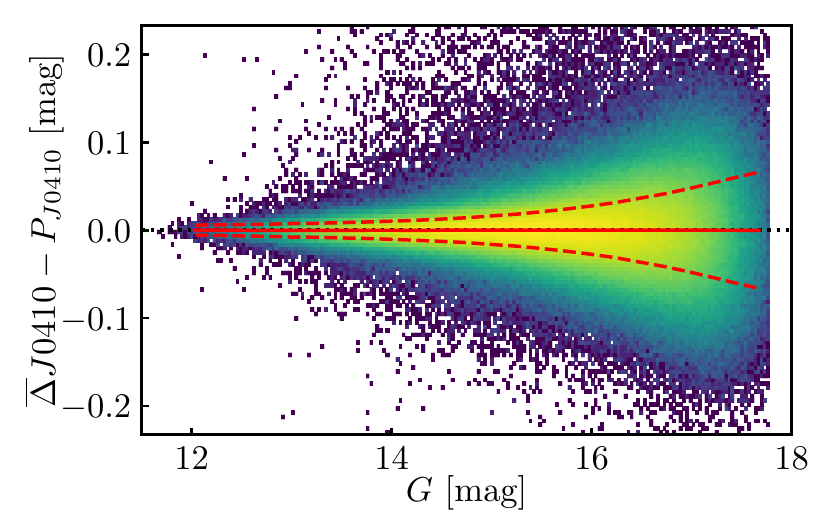}}
\resizebox{0.245\hsize}{!}{\includegraphics{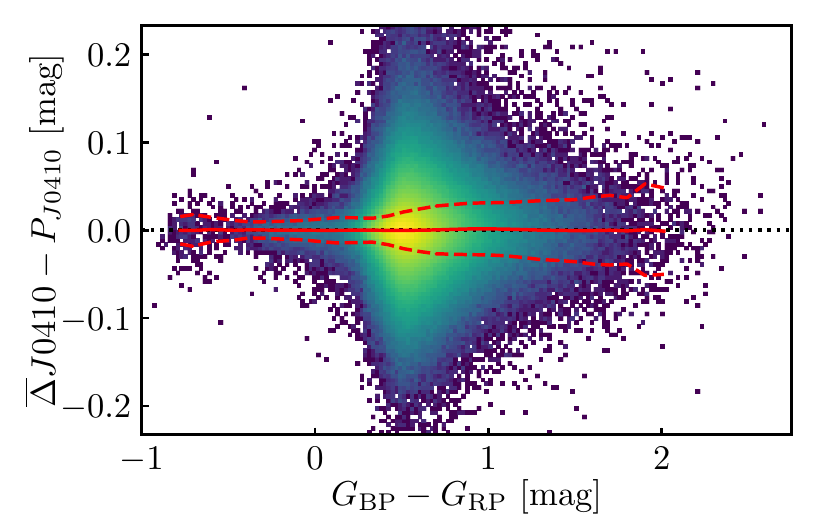}}\\
\resizebox{0.245\hsize}{!}{\includegraphics{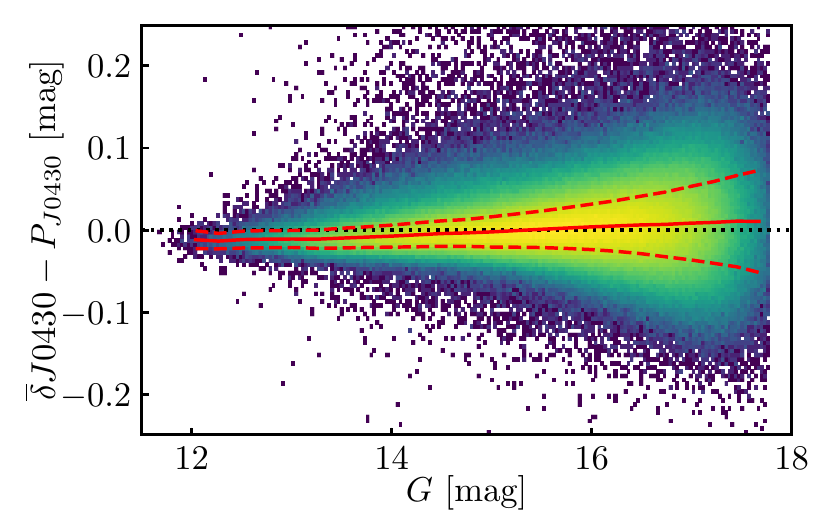}}
\resizebox{0.245\hsize}{!}{\includegraphics{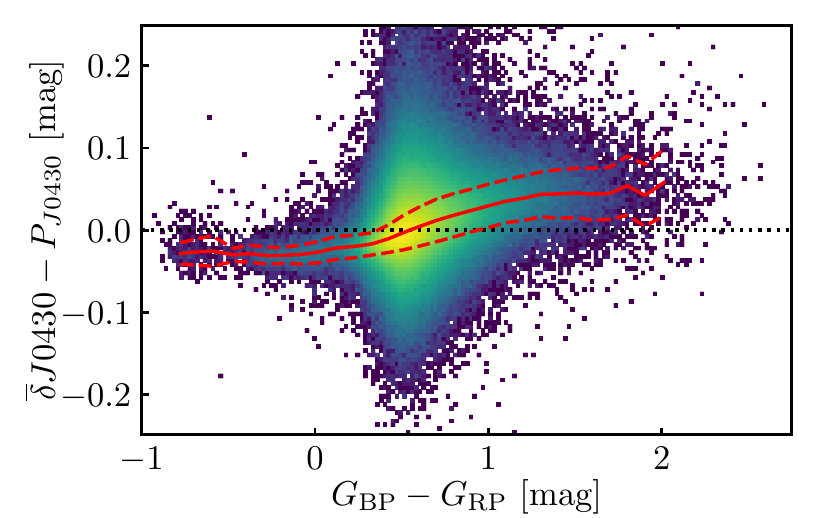}}
\resizebox{0.245\hsize}{!}{\includegraphics{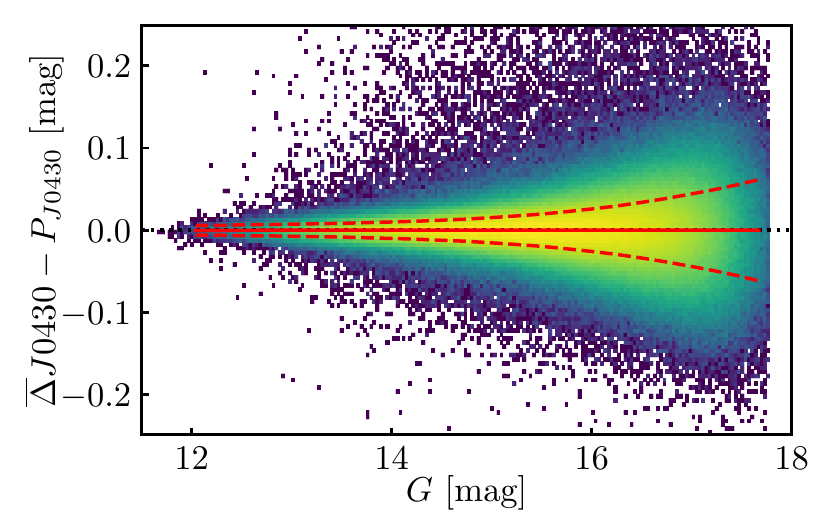}}
\resizebox{0.245\hsize}{!}{\includegraphics{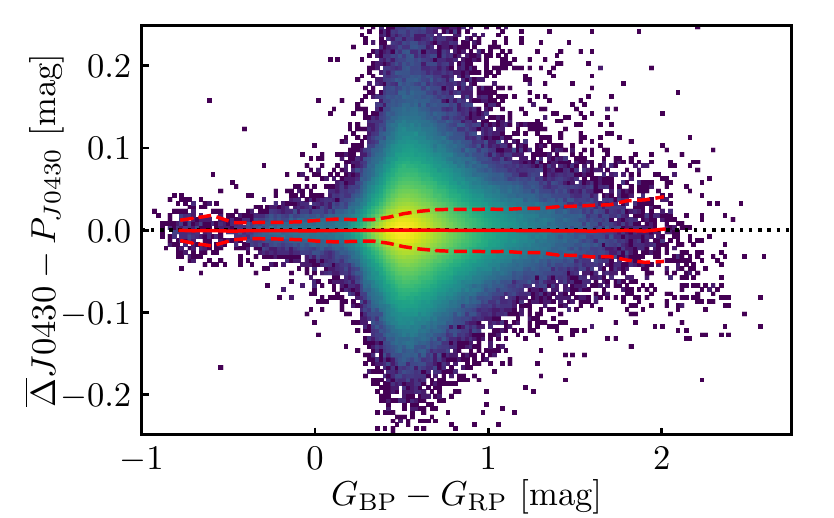}}\\
\resizebox{0.245\hsize}{!}{\includegraphics{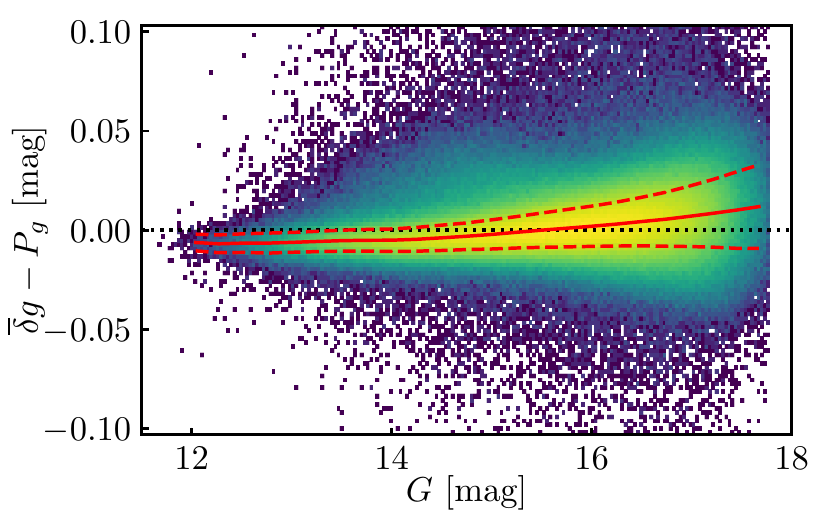}}
\resizebox{0.245\hsize}{!}{\includegraphics{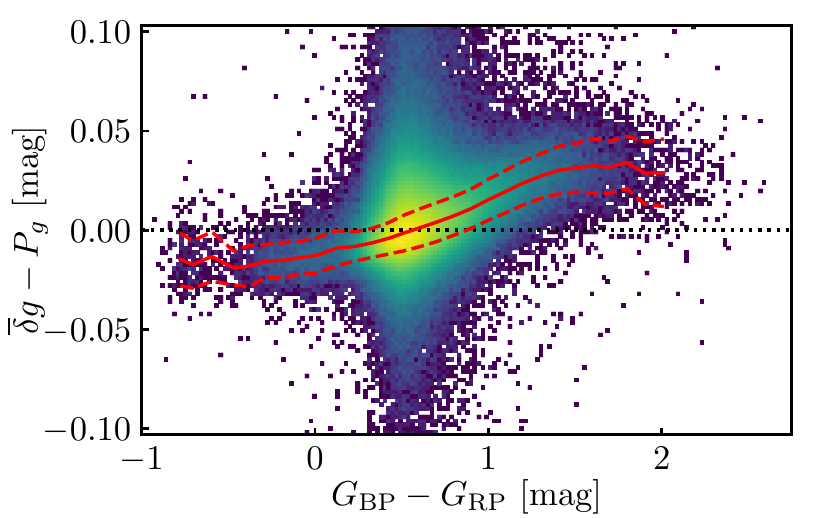}}
\resizebox{0.245\hsize}{!}{\includegraphics{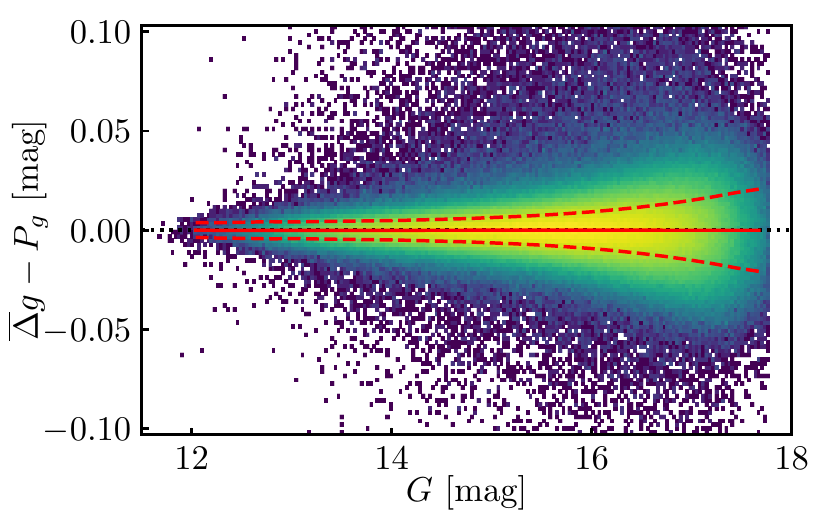}}
\resizebox{0.245\hsize}{!}{\includegraphics{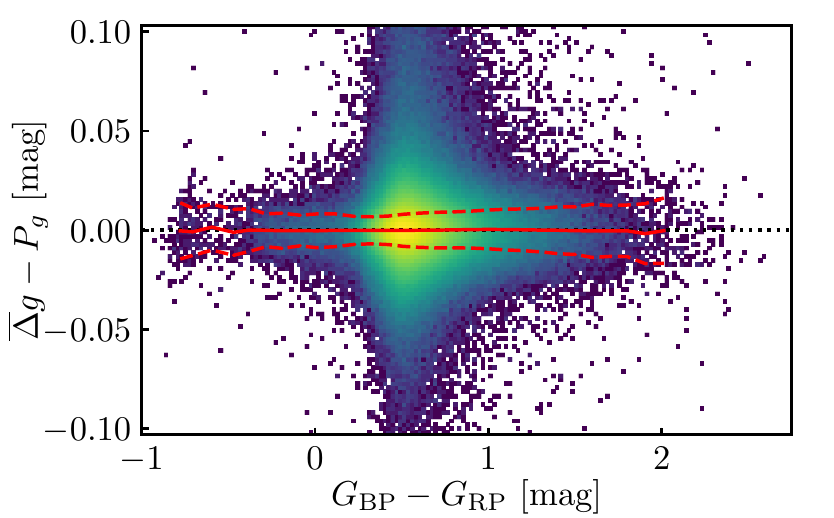}}\\
\resizebox{0.245\hsize}{!}{\includegraphics{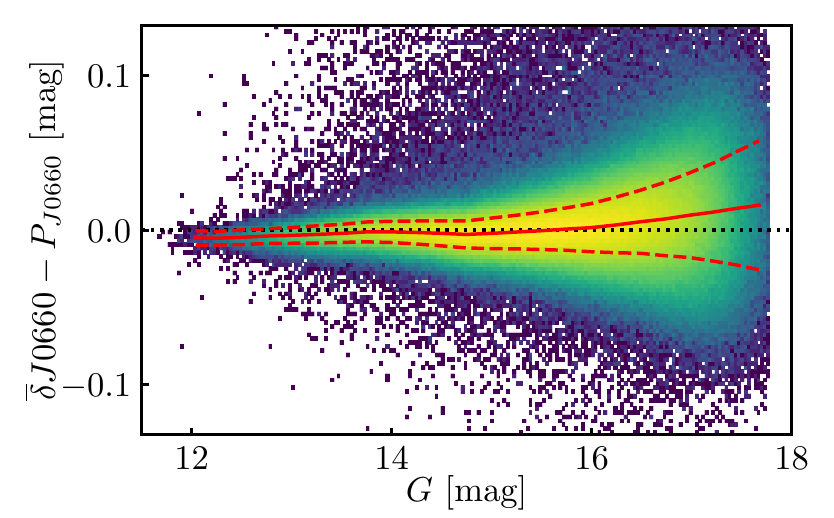}}
\resizebox{0.245\hsize}{!}{\includegraphics{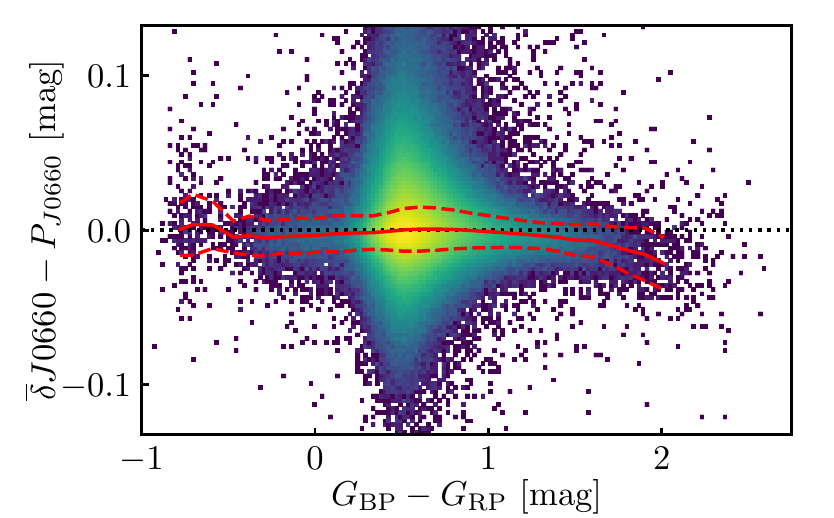}}
\resizebox{0.245\hsize}{!}{\includegraphics{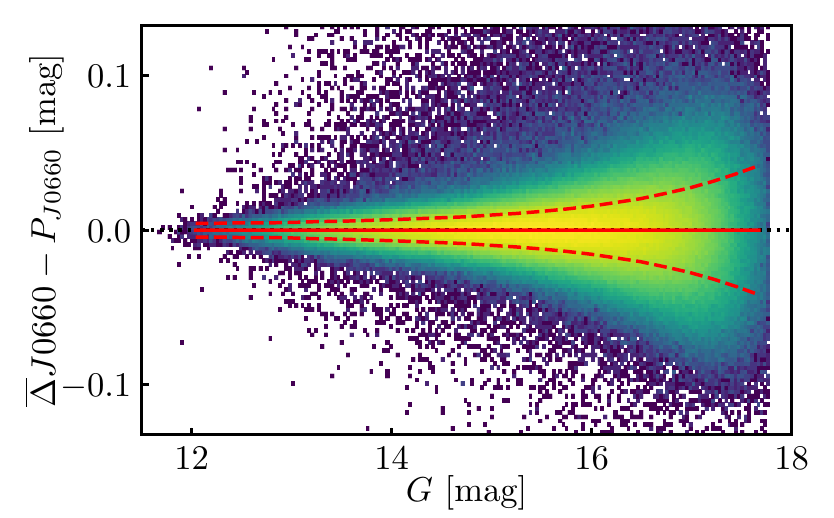}}
\resizebox{0.245\hsize}{!}{\includegraphics{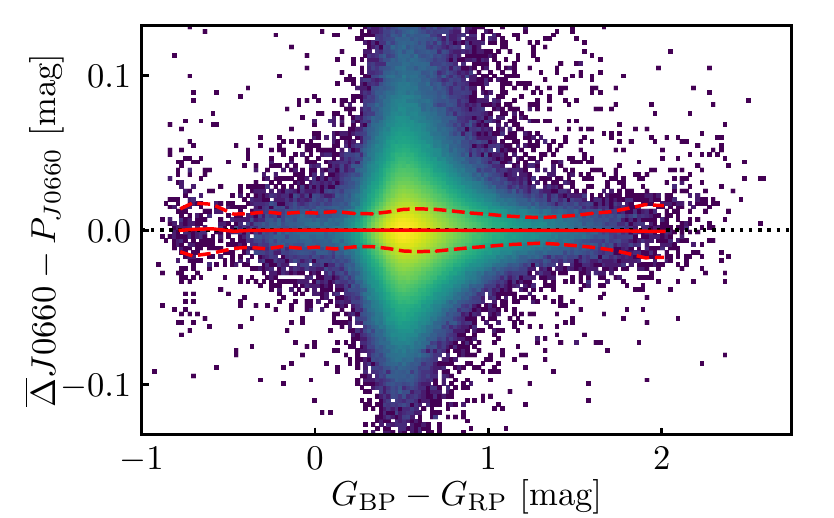}}\\
\resizebox{0.245\hsize}{!}{\includegraphics{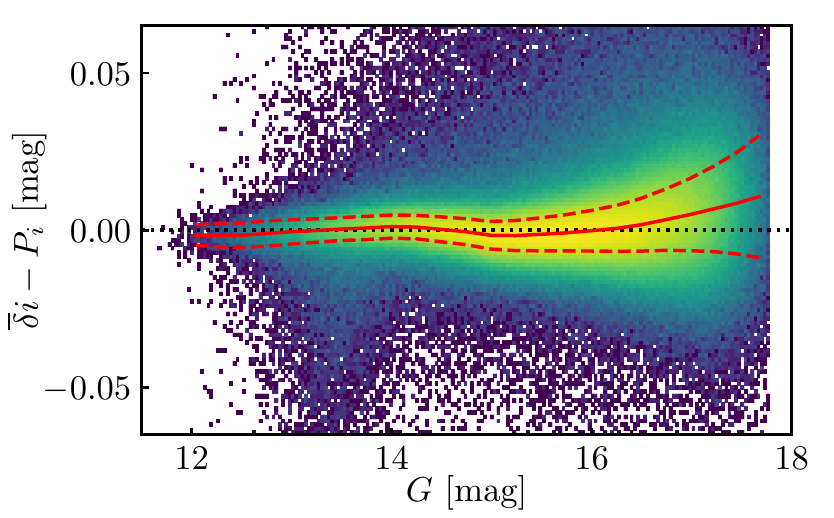}}
\resizebox{0.245\hsize}{!}{\includegraphics{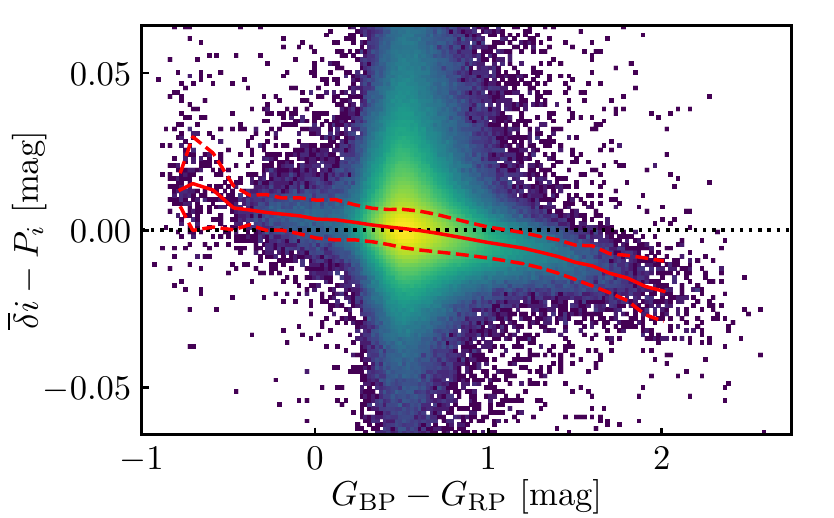}}
\resizebox{0.245\hsize}{!}{\includegraphics{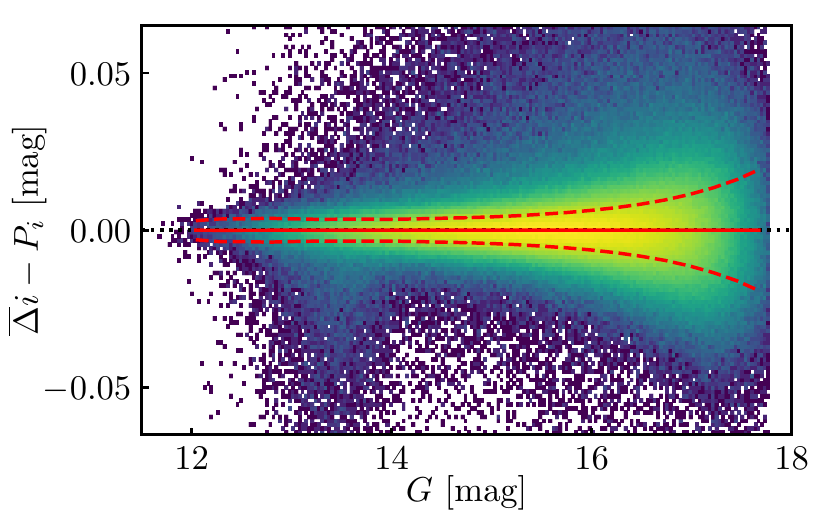}}
\resizebox{0.245\hsize}{!}{\includegraphics{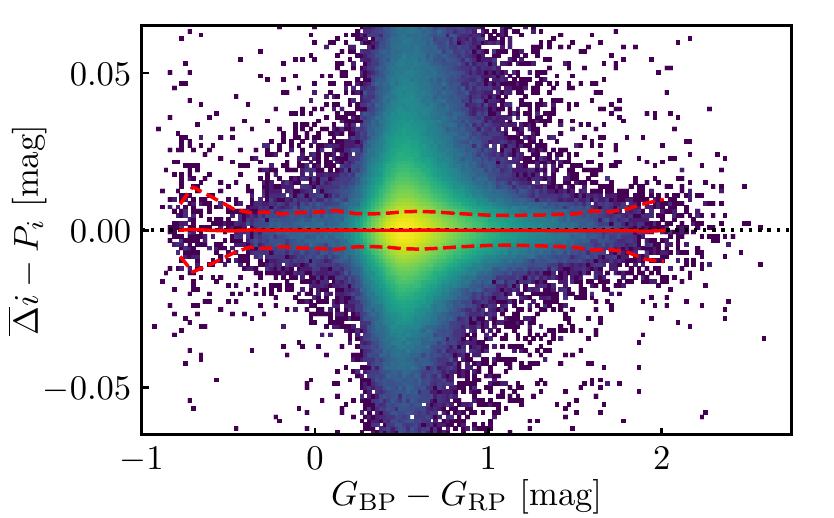}}
\caption{Residuals between the synthetic photometry from {\it Gaia} BP/RP spectra and J-PLUS photometry in the the $J0378$, $J0395$, $J0410$, $J0430$, $g$, $J0660$, and $i$ passbands from top to bottom. The panels from left to right show the residuals as a function of the $G$ magnitude and the $\bprp$ color without the transformation terms $T_{r}^{\rm mag}$ and $T_{r}^{\rm col}$, and as a function of the $G$ magnitude and the $\bprp$ color after applying them. The color scale depict the number density of sources using a logarithm scale, with nearly $1.5$ million sources shown. In all the panels, the solid-red line represents the the median of the residuals and the dashed lines its one $\sigma$ dispersion. 
} 
\label{fig:Tfiltro1}
\end{figure*}

\begin{figure*}[t]
\centering
\resizebox{0.245\hsize}{!}{\includegraphics{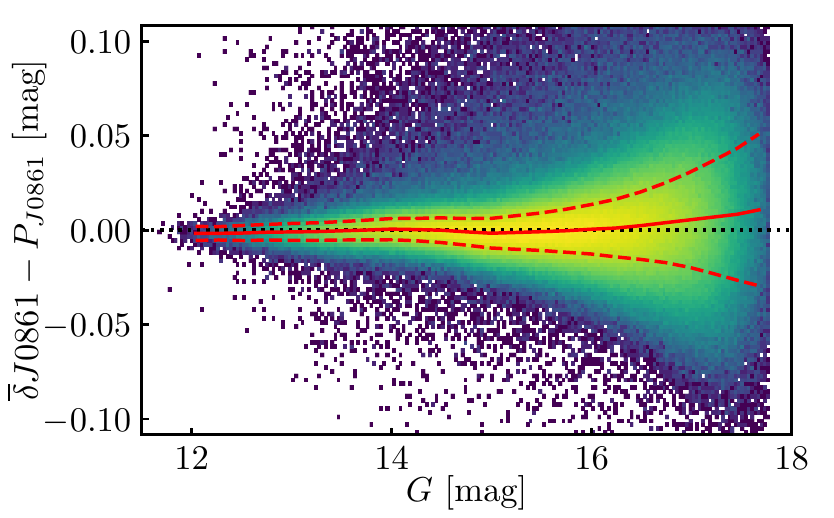}}
\resizebox{0.245\hsize}{!}{\includegraphics{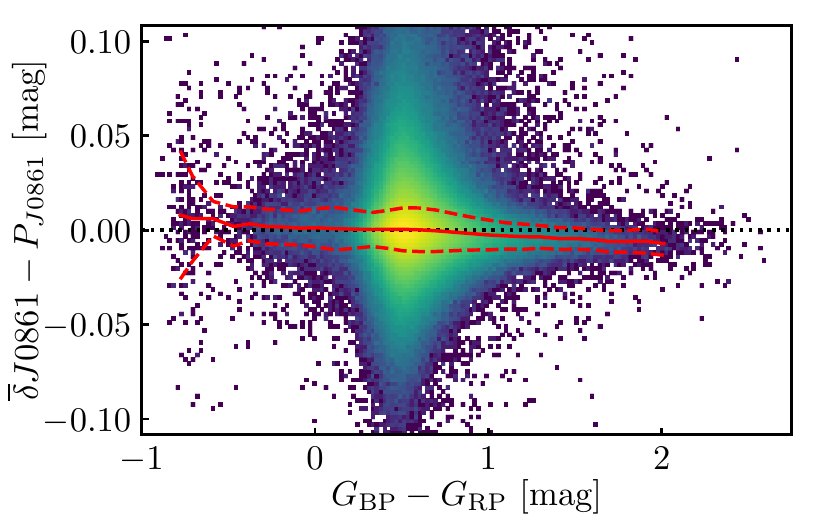}}
\resizebox{0.245\hsize}{!}{\includegraphics{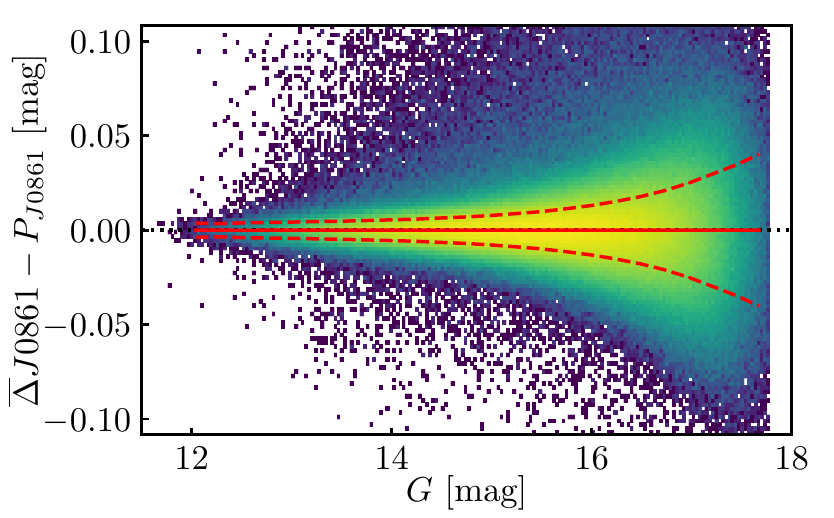}}
\resizebox{0.245\hsize}{!}{\includegraphics{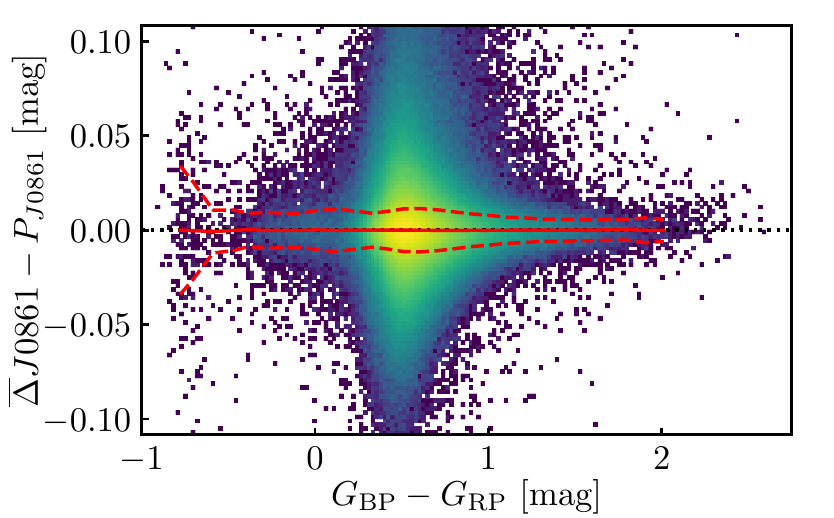}}\\
\resizebox{0.245\hsize}{!}{\includegraphics{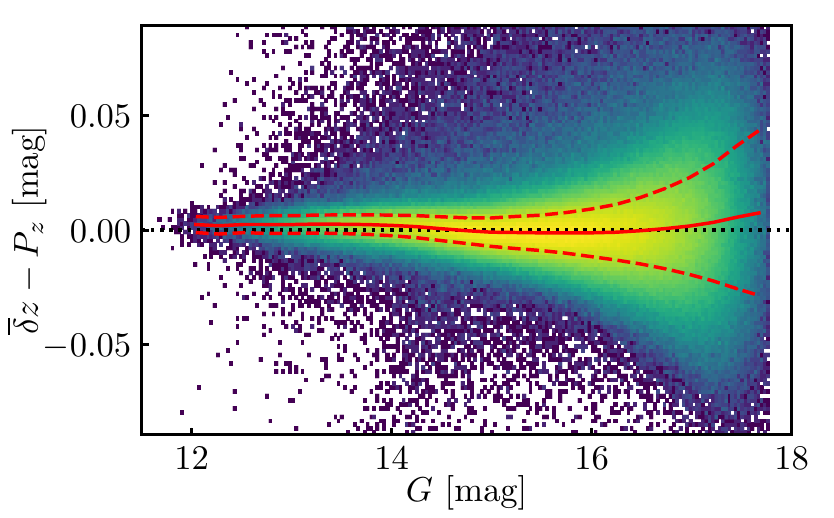}}
\resizebox{0.245\hsize}{!}{\includegraphics{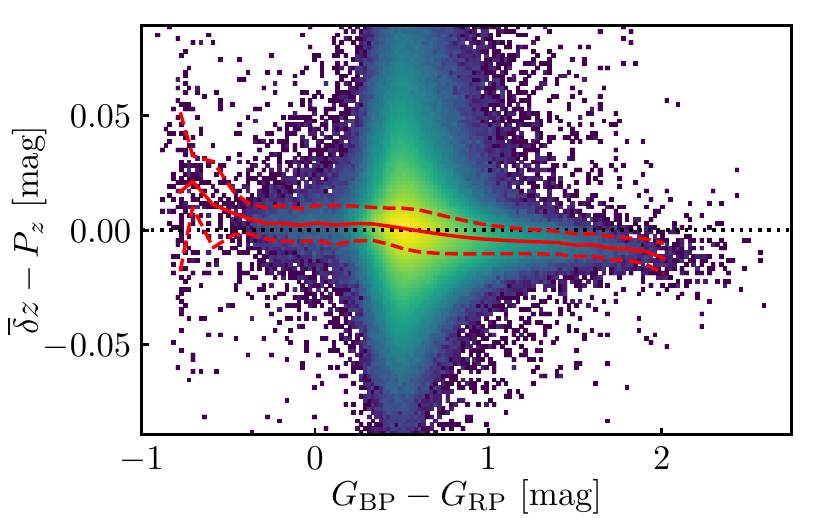}}
\resizebox{0.245\hsize}{!}{\includegraphics{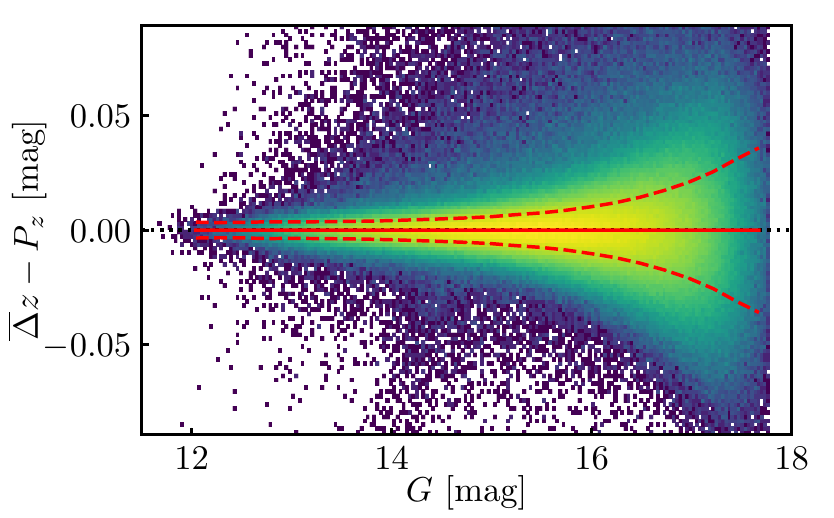}}
\resizebox{0.245\hsize}{!}{\includegraphics{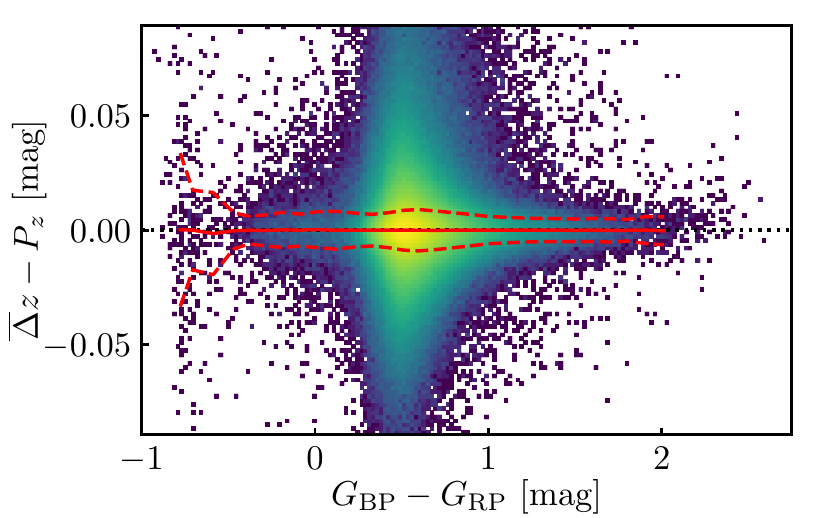}}
\caption{Residuals in the $J0861$ and $z$ passbands, following Fig.~\ref{fig:Tfiltro1}.
} 
\label{fig:Tfiltro2}
\end{figure*}

The original and corrected residuals between {\it Gaia} BP/RP synthetic photometry and J-PLUS instrumental magnitudes as a function of $G$ magnitude and $\bprp$ color for the J-PLUS passbands $J0378$, $J0395$, $J0410$, $J0430$, $g$, $r$, $J0660$, $i$, $J0861$, and $z$ are presented in Figs.~\ref{fig:Tfiltro1} and \ref{fig:Tfiltro2}.

\section{White dwarf locus models}\label{app:wdlocus}
\begin{figure*}[t]
\centering
\resizebox{0.49\hsize}{!}{\includegraphics{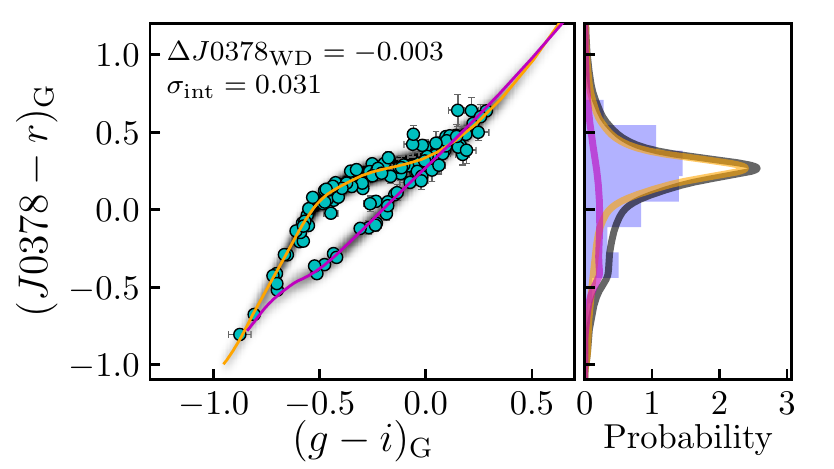}}
\resizebox{0.49\hsize}{!}{\includegraphics{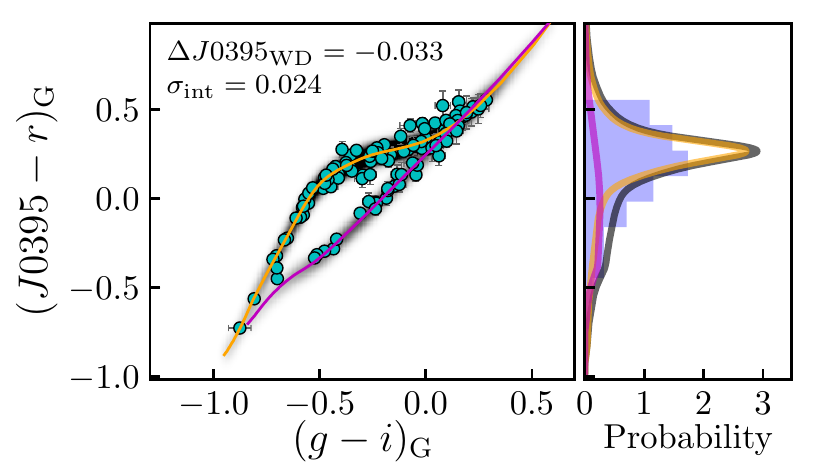}}

\resizebox{0.49\hsize}{!}{\includegraphics{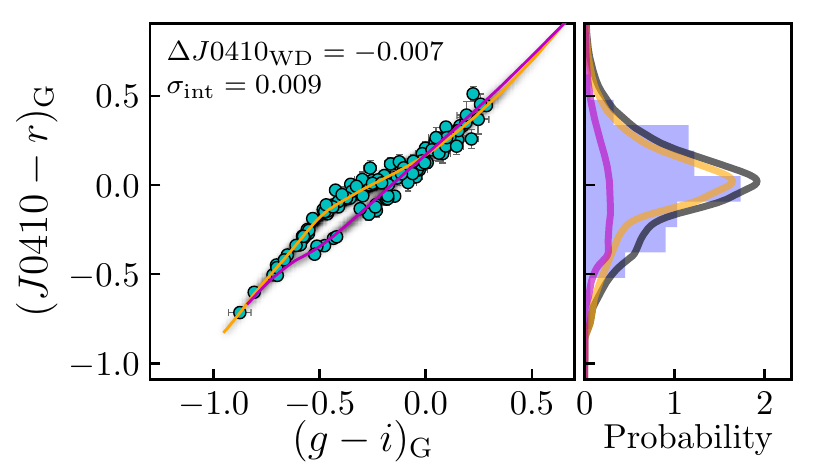}}
\resizebox{0.49\hsize}{!}{\includegraphics{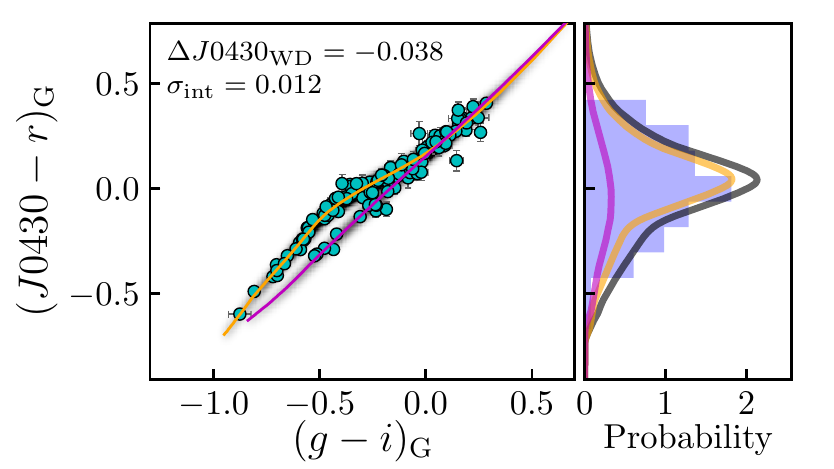}}

\resizebox{0.49\hsize}{!}{\includegraphics{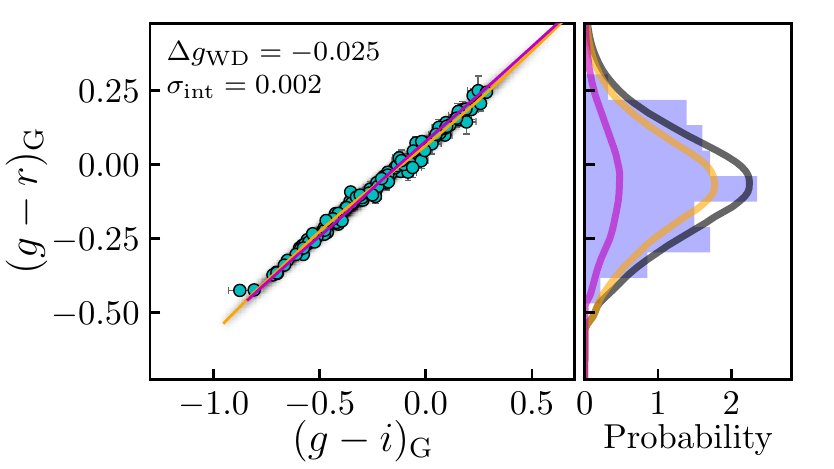}}
\resizebox{0.49\hsize}{!}{\includegraphics{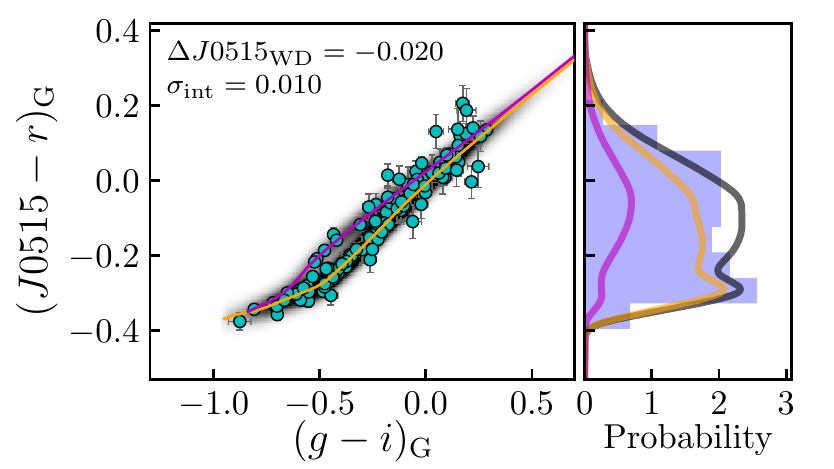}}
\caption{Similar to Fig.~\ref{fig:wdlocus_1}, but for $\mathcal{X}$ = $J0378$, $J0395$, $J0410$, $J0430$, $g$, and $J0515$ passbands. We omit the $(g-i)_{\rm G}$ projection because it is shared by all the panels.}
\label{fig:wdlocus_2}
\end{figure*}

\begin{figure*}[t]
\centering
\resizebox{0.49\hsize}{!}{\includegraphics{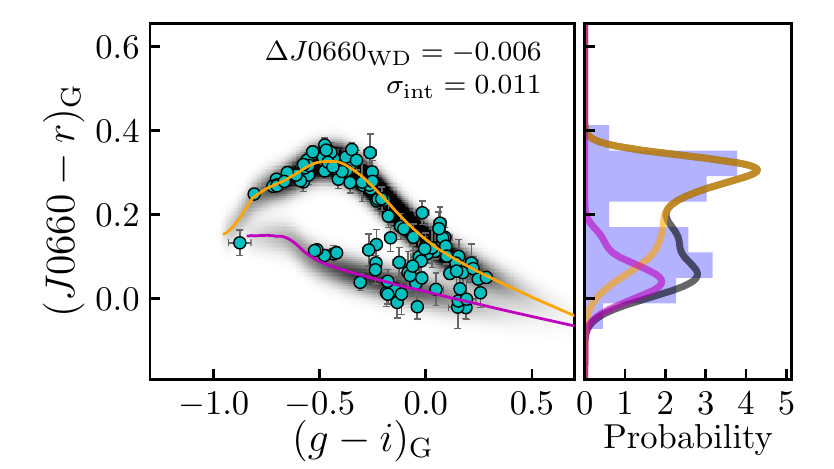}}
\resizebox{0.49\hsize}{!}{\includegraphics{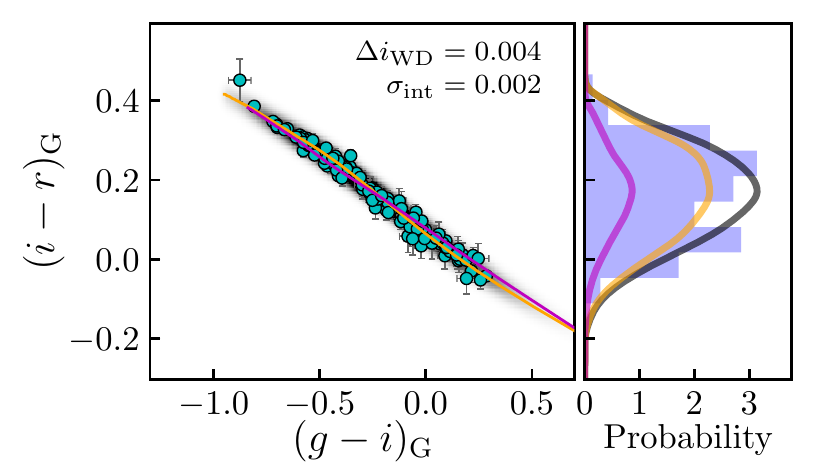}}

\resizebox{0.49\hsize}{!}{\includegraphics{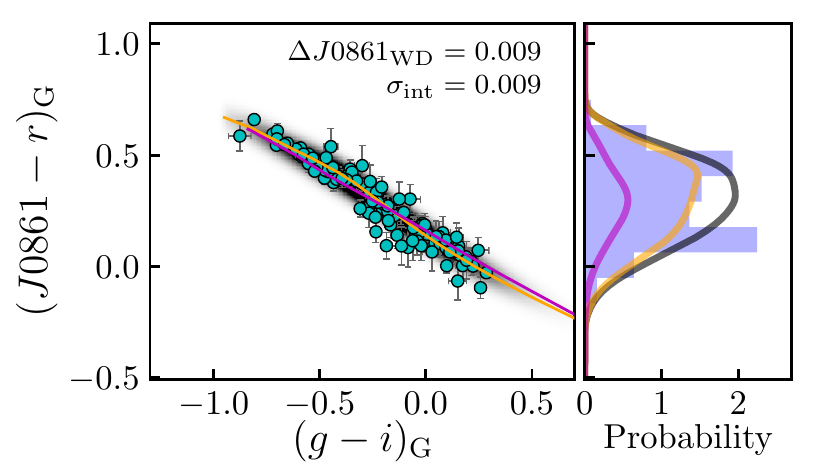}}
\resizebox{0.49\hsize}{!}{\includegraphics{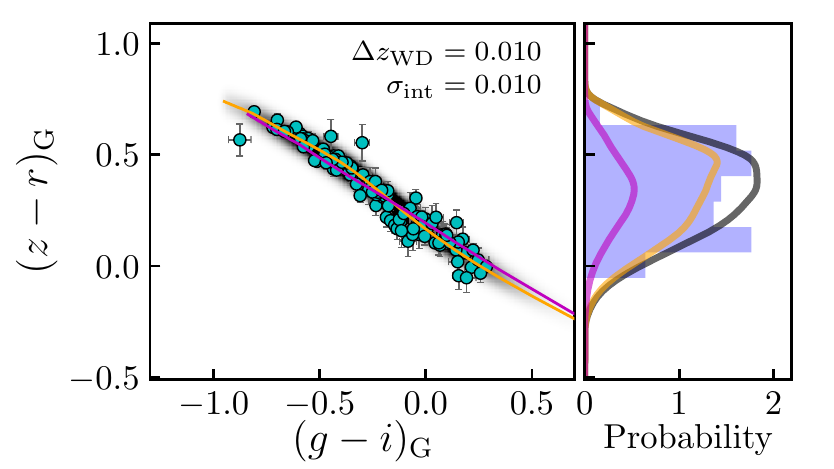}}
\caption{Similar to Fig.~\ref{fig:wdlocus_1}, but for $\mathcal{X}$ = $J0660$, $i$, $J0861$, and $z$ passbands. We omit the $(g-i)_{\rm G}$ projection because it is shared by all the panels.}
\label{fig:wdlocus_3}
\end{figure*}

The results for the joint Bayesian modeling of the white dwarf locus presented in Sec.~\ref{sec:wdlocus} are shown in Figs.~\ref{fig:wdlocus_2} and \ref{fig:wdlocus_3}. 

\end{appendix}

\end{document}